\documentclass[preprint,12pt]{elsarticle}
\usepackage{psfrag}
\usepackage[T1]{fontenc}
\usepackage{epsfig}
\usepackage{times}
\usepackage{amsmath}
\usepackage{amssymb}
\usepackage{graphicx}
\usepackage{wasysym}
\usepackage{textcomp}
\input epsf.sty
\newcommand{\be}{\begin{equation}}
\newcommand{\ee}{\end{equation}}
\newcommand{\ba}{\begin{eqnarray}}
\newcommand{\ea}{\end{eqnarray}}
\newcommand{\bi}{\begin{itemize}}
\newcommand{\ei}{\end{itemize}}

\renewcommand{\vec}[1]{{\bf #1}}

\def\lsi{\raise0.3ex\hbox{$<$\kern-0.75em\raise-1.1ex\hbox{$\sim$}}}
\def\gsi{\raise0.3ex\hbox{$>$\kern-0.75em\raise-1.1ex\hbox{$\sim$}}}

\newcommand{\BE}{\begin{eqnarray}}
\newcommand{\EE}{\end{eqnarray}}
\newcommand{\BDM}{\begin{displaymath}}
\newcommand{\EDM}{\end{displaymath}}

\makeatletter \@addtoreset{equation}{section} \makeatother

\begin{document}

\begin{frontmatter}

\title{Quantum Searches in a Hard 2SAT Ensemble}

\author[jsc]{T. Neuhaus}

\address[jsc]{J\"ulich Supercomputing Centre,Forschungszentrum J\"ulich, D-52425 J\"ulich, Germany}

\begin{abstract}

Using a recently constructed ensemble 
of hard 2SAT realizations,
that has a unique ground-state 
we calculate for the quantized theory the median gap correlation
length values $\xi_{GAP}$ along the direction of the
quantum adiabatic control parameter $\lambda$.
We use quantum annealing (QA) with transverse
field and a linear time schedule in the 
adiabatic control parameter $\lambda$. 
The gap correlation length 
diverges exponentially $\xi_{\rm GAP}
\propto {\rm exp} [+r_{\rm GAP}N]$ in the median with a 
rate constant $r_{\rm GAP}=0.553(6)$, while the run time 
diverges exponentially $\tau_{\rm QA}
\propto {\rm exp} [+r_{\rm QA}N]$ with $r_{\rm QA}=1.184(16)$. 
Simulated classical annealing (SA) exhibits a run time rate 
constant $r_{\rm SA}=0.340(5)$ that is small and thus
finds ground-states exponentially faster 
than QA. There are no quantum speedups in ground state searches
on constant energy surfaces that have exponentially large volume.
We also determine gap correlation length 
distribution functions 
$P(\xi_{\rm GAP})d\xi_{\rm GAP} \approx W_k$ over the ensemble
that at $N=18$ are close to Weibull
functions $W_k$ with $k \approx 1.2$ i.e., the problems show
thin catastrophic tails in $\xi_{\rm GAP}$. The inferred
success probability distribution functions of the quantum annealer
turn out to be bimodal.

\end{abstract}

\end{frontmatter}

\section{Introduction}

The purpose of the paper is to elaborate on 
the aspects of search 
efficiency's in quantum ground 
state searches in specific models, toy models for that matter. 
We expand our theoretical knowledge
and consider a mathematical theory, namely 
two-satisfiability (2SAT) that within Cook's 
classification \cite{Cook} of mathematical 
complexity lies in $\cal{P}$ for sure. 
In that case mathematical rules of calculation 
are known, that solve any problem as 
a function of the  number of degrees of freedom $N$ just 
in polynomial time. 
To be specific: Any 2SAT problem is mapped to a directed 
graph of implications, which then is 
analyzed for logical collisions: 
A task that algorithmically 
can be accomplished in polynomial compute time and knowing
all collisions actually determines the ground-state
in satisfiable instances. A comparable situation is 
encountered in the planar $2d$
Ising glass without external field, where also algorithms
exist that find ground-states in polynomial time, see 
\cite{Hartmann_02011}. 

The current work complements a recent work on 
synthetic ensembles of hard
2SAT \cite{Neuhaus_on_hard_problems} problems, which 
in vicinity of the satisfiability 
threshold of random 2SAT
at $\alpha=M/N = 1$ \cite{Borgs_et_al_02001}, and 
for satisfiable instances 
are designed to hide a {\rm unique} 
ground state of the 2SAT Hamiltonian
$H_{2SAT}$ at energy $E=0$ from a vast number of non-solutions
at the energy gap $E=1$. 
Biasing toward hard problems follows
an idea of Znidaric \cite{Znidaric_2005_3sat_problems,Znidaric_2005} 
for 3-SAT, which is worked 
out further in \cite{Neuhaus_on_hard_problems} 
for K-SAT with $2 \le K \le 6$.
The ensemble of hard 2SAT realizations
has a exponential divergence
\begin{equation}
<\Omega(E=1)>_{\rm 2SAT,HARD} \propto {\rm} e^{+r_{\rm DOS}N}
\label{dos_singularity}
\end{equation}
for the density of states function in 
the median of the ensemble at energy $E=1$.
The numeric value of the rate constant for 2SAT
$r_{\rm DOS}=0.329(2)$
equals the entropy density $s_1={\rm ln}\Omega(E=1)/N$
at energy one and  
is a numeric model parameter without physical meaning.
It parametrizes a doubling of the number of energy one configurations if 
about two new spins are added to the theory.
Thus any stochastic search for the
single ground state at $E=0$ within the energy one surface 
is expected to have serious problems. We expect this to be 
true for {\it simulated annealing} (SA) 
\cite{Kirkpatrick_Gelatt_Vecchi_1983} as well
as well for {\it quantum annealing} (QA) 
\cite{Apolloni_1989,quantum_chemist_1993,quantum_chemist_1994,Nishimori_1998,Morita_Nishimoro_2008,Ohzeki_Nishimori_02011}
and we consider quantum adiabatic 
computations \cite{Farhi_02001} 
to be a version of quantum annealing.
It is important to note that our problems
are not generated by heuristics but have a
proper partition function representation.

Our study is part of a research effort
that studies quantum annealing for the
Ising Hamiltonian $H_{\rm Ising}$ in 
terms of local Pauli matrices $\vec{\sigma}_i$
and two-point spin couplings at finite magnetic field $h_i$:
\begin{equation}
H_{\rm Ising }=-\sum_{<ij>}J_{ij} \sigma_i^z \sigma_j^z
+\sum_i h_i  \sigma_i^z 
- \Gamma \sum_i \sigma_i^x,
\label{ising_hamiltonian}
\end{equation}
where the first two terms denote the problem Hamiltonian. 
The particular 2SAT Hamiltonian 
$H_{2SAT}$ of this work, and after a trivial reduction step 
from logical literals to Ising spins, has in fact the exact form 
of the problem part in eq.(\ref{ising_hamiltonian}). 
The ensemble of problems has non-vanishing magnetic 
fields $h_i \ne 0$, frustration via sign alternating
two-point couplings $\pm J_{i,j}$, bounded couplings $J$ and $h$,
a finite classical energy gap of unity and is defined on a 
non-planar connectivity graph at rather sparse connectivity: there are only
$N+1$ two-point couplings at spin number $N$. 
If anything: The satisfiability problem as studied here represent
a particular hard and simple synthetic problem class for physical annealers 
within the set of all realizations of the form $H_{\rm Ising}$.

Within the last decade it has been argued 
\cite{Nishimori_1998,Farhi_02001,Santoro_et_al_2002,Santoro_et_al_2002_long} 
that quantum annealing in general and in particular for 
frustrated $Z(2)$ Ising degrees of freedom can show faster convergence
to the ground state of the problem Hamiltonian 
than classical annealing, see also the extensive reviews 
\cite{tosatti_review_2006,Chakrabarti_review_2008}.
We think however that as of today it is fair to say that
for Hamiltonian's of the form eq.(\ref{ising_hamiltonian})
there is no convincing demonstration, that any 
frustrated Ising theory on non-planar connectivity graphs 
shows in fact a speedup in quantum annealing 
over classical annealing in the limit of large $N$, a view 
that is also shared in 
\cite{Altshuler_02010} and even is experimentally
studied \cite{Troyer_02013,Troyer_02014} on an 
existing annealing device with negative finding. 
Recent quantum annealing numerical studies
in the 2D glass enhance the skepticism  \cite{Troyer_02014_SA_vs_QA}. 
The findings of this work also do not 
suggest that quantum speedups
are easy to obtain.
In fact: The Ising model realizations
of this work exhibit a clear and indisputable 
exponential speedup of simulated annealing
over quantum annealing. We note that additional
studies in other satisfiability theories like 3-XORSAT 
\cite{PY10,PY11,PY12} and
3-SAT \cite{Battaglia_Santoro_and_Tosatti,Znidaric_2005,Neuhaus_2011} 
showed, that quantum annealing does not solve these 
theories efficiently either. 

Finally an important motivation of the current work
is the presentation of precise 
numerical data for the energy gap at the quantum transition, which 
can be confronted to other 
approaches that deal with quantum search 
complexity. These include the semi-perturbative evaluation
of the quantum energy gap \cite{Amin_02012}, Quantum 
Monte Carlo annealing 
\cite{Nishimori_1998,Brooke_1999,Santoro_et_al_2002,Neuhaus_quantum_annealing_02011} 
as well as annealing studies under Schr\"odinger wave 
function dynamics \cite{Hans_d_R_1987,Katzgraber_02009}. 
We expect that algorithmic studies e.g. studies of {\it simulated}
quantum annealing can profit as the problem ensemble here 
has small number of spins, while at the same time 
the maximal quantum gap correlation length 
is as large as $\xi_{\rm GAP} \approx 4730$ for a theory with only 
$N=18$ spins.



\section{Theory, Observables and Basic Observations on Annealing}

\subsection{Theory}

We consider 2SAT problems
with $i=1,\ldots,N$ Boolean variables $x_i=0,1$ and $M$ clauses.
In 2-satisfiability one tries to
find a truth assignment to the variables $x_i$ that makes 
the conjunctive normal form
\begin{equation}
{\cal F}=( L_{1,1} \lor L_{1,2})
\land (   L_{2,1} \lor L_{2,2})
\land ...
\land (L_{M,1} \lor L_{M,2})
\end{equation}
of $M$ clauses true, if that is possible in which case
a problem is satisfiable. 
There are $2 \times M$ 
literals $L_{j,k}$ with $j=1,\ldots,M$ and $k=1,2$. Each literal
being a variable $x_i$ or its negation $\overline{x}_i$ and if 
the assignment of variables to literals and possible negations
are chosen at random the issue is non-trivial. 
The problem is equivalent to finding ground-states at 
energy $E=0$ of the Hamiltonian
\begin{equation}
H_{\rm 2SAT}= \sum_{j=1}^{M}
h^{(2)}(\epsilon_{j,1}s(L_{j,1}),\epsilon_{j,2}s(L_{j,2}))
\label{2SAT_hamiltonian}
\end{equation}
where
\begin{equation}
h^{(2)}(s_l,s_m)=\frac{s_l-1}{2}\frac{s_m-1}{2},
\end{equation}
with $l,m\in\{1,\ldots,N\}$.
The matrix elements $\epsilon_{j,k}$ with $j=1,\ldots,M$ and $k=1,2$ 
take the value $-1$ if a variable is negated and $+1$ otherwise
and $s(L_{j,k})$ denotes a map from literals $L_{j,k}$ to 
Ising spins $s_i=\pm 1$ with $i=1,...,N$.
The Hamiltonian $H_{\rm 2SAT}$ has an equally 
spaced discrete energy spectrum with a ground-state energy $E=0$
for satisfiable instances and a first excited energy gap of 
value one and, we will only consider 
satisfiable instances in this work $\Omega(E=0) \ge 1$.
Physically the Hamiltonian describes
an infinite dimensional Ising 
spin glass of the Edwards Anderson type 
with two values of frustration $J=\pm {1 \over 4}$ in a
finite magnetic random field. The coordination number can 
be tuned by $M/N$, the 
connectivity is free and thus planar as well as non-planar connectivity 
graphs can be involved.

\begin{figure}[htb*]
\centering{
\includegraphics[angle=-90,width=8.5cm]{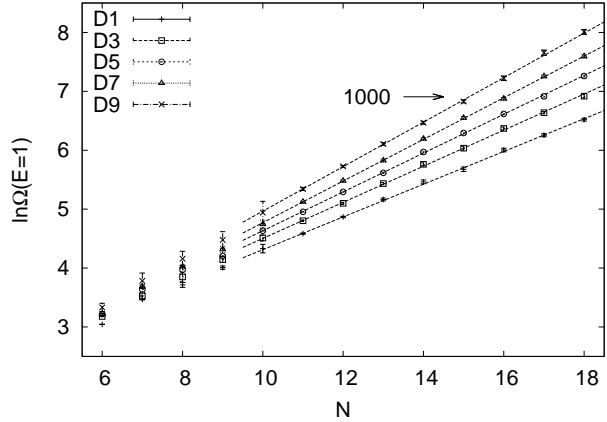}
\vspace*{0.0cm}
\caption{Density of states ${\rm ln}\Omega(E=1)$ in Deciles
for the hard 2SAT problem ensemble as a function of $N$. 
The median is denoted by $D5$. The straight
lines demonstrate an exponential increase of $\Omega(E=1)$ 
with rate constants $r_{\rm DOS}$ 
as given in Table 1). We display the Deciles $Dn$ with $n=1,3,5,7,9$.}
}
\end{figure}
~~~~~~~~~~~~~~~~~~~~~~~~~~~~~~~~~~~~~~~~~
\begin{table}[htb*]
\centering{
\begin{tabular}{|c|c|c|c|c|c|} 
\hline      
${\rm Decile}$ & $N_{\rm min}$ & $r_{\rm DOS}$ & $\chi^2_{dof}$  \\
\hline\hline
  1 & 10 & 0.277 ( 2 ) & 0.57 \\ 
  3 & 10 & 0.306 ( 2 ) & 0.89 \\ 
  5 & 10 & {\bf 0.329 ( 2 )} & 0.36 \\ 
  7 & 10 & 0.353 ( 2 ) & 0.22 \\ 
  9 & 10 & 0.377 ( 2 ) & 0.39 \\  
\hline\hline
\end{tabular}
\caption{Fit values of slopes $r_{\rm DOS}$ 
for straight line fits in Fig. 1).
The fits include data at $N$ with  
$N \ge N_{\rm min}$ and the $\chi^2_{dof}$-values of the fits with the form 
eq.(\ref{dos_singularity}) is given. $D5$ denotes the median value and 
is highlighted by fat symbols.}
}
\end{table}

We employ special 2SAT problems with 
unique ground-state i.e., density of states function $\Omega(E=0)=1$
from a recently constructed synthetic 
ensemble \cite{Neuhaus_on_hard_problems}.
The hard 2SAT problem ensemble is designed to hide the
ground state at energy
$E=0$ from a large number of non-solutions at energy $E=1$ and has
a clause-to-variable ratio 
$\alpha\equiv M/N=(N+1)/N \approx 1$ for $3 \le N \le 18$, that
asymptotically for large $N$ approaches the satisfiability threshold 
of random 2SAT. The situation is depicted in Fig. 1)
where a selected set of quantiles i.e., Deciles of the 
quantity ${\rm ln}\Omega(E=1)$ on the cumulative problem 
distribution is displayed as a function of $N$. It can be witnessed
that $\Omega(E=1)$ diverges exponentially in $N$, the straight lines
in Fig. 1), in accord with eq.(\ref{dos_singularity}). 
The fifth decile (D5) is the median and a fit with $N \ge 9$
and the form  eq.(\ref{dos_singularity}) readily yields 
the rate constant $r_{\rm DOS}=0.329(2)$ at a $\chi^2_{dof}=0.36$ 
for the fit, see also Table 1). 
There is a certain spread of the energy one entropy density
$s_1=r_{\rm DOS}$, which for the first decile D1 ranges
from $s_1=0.278$ up to $s_1=0.377$ at D9, which is a characteristic
of underlying catastrophic statistics. The bias
toward the hard 2SAT problem set was implemented in 
\cite{Neuhaus_on_hard_problems} via a systematic 
Monte Carlo study of a generating partition function
\be
\Gamma (\mu) ~~~=~~~\sum_{\rm Random~2SAT } ~~~e^{W_{\rm MUCA}(\mu)}~~~\delta^{(1)} [\mu -\Omega(E=0)]
\label{generating_function}
\ee
at fluctuating $\mu$-value, where 
$W_{\rm MUCA}$ is a multicanonical weight \cite{Muca}.
The Monte Carlo process then generates at $\mu=1$ the hard 
2SAT problem ensemble. Unfortunately and as of 
today it was not possible to generate problems at larger values 
of $N$ than $N = 18$. 
The reason being: The Monte Carlo moves on 
random 2SAT require calculations of $\Omega(E=0)$ which by itself
creates an exponentially large compute barrier. We have spend some time
to alleviate the situation but in essence were not successful.

\subsection{Quantum Annealing}

We find the unique ground state of the hard 2SAT instances 
via quantum annealing 
\cite{Apolloni_1989,quantum_chemist_1993,quantum_chemist_1994,Nishimori_1998,Morita_Nishimoro_2008,Ohzeki_Nishimori_02011}
and use in particular the quantum adiabatic computations \cite{Farhi_02001} .
Conventional adiabatic quantum computation (QAC) 
assumes a linear interpolation between the problem 
Hamiltonian $H_{\rm 2SAT}$ of eq.(\ref{2SAT_hamiltonian}) and 
a non-commuting driver Hamiltonian 
$H_D=\sum_i\sigma_i^x$, the transverse field term
where $\sigma_i^x$ is the $x$-component Pauli matrix. The linear
interpolation is:
\begin{equation}
H_{QAC}(\lambda)=  (1-\lambda)H_D + \lambda H_{\rm 2SAT}.
\label{HQAC}
\end{equation}
The quantum adiabatic parameter is defined on the compact
interval $0 \le \lambda \le 1$ and in a physical annealing
device becomes a smoothly varying function $\lambda(t)$ of 
physical time $t$ such that $\lambda (t=0)=0$ 
and $\lambda(t={\cal T})=1$.
The time $t={\cal T}$ denotes the annealing time of single annealing 
trajectories and, as well as on devices and in theory 
trajectories are repeated many times yielding
a success probability $P_{\rm Success}({\cal T})$ with 
$0 < P_{\rm Success} \le  1$ 
for successful ground-state searches. 
Clearly a single annealing run 
of possible short duration ${\cal T}$ may miss the ground-state
but the repetition of many results into finite success.
Combinatorics then tells us that 
\be
\hat{\tau}_{\rm QA} (P_{\rm Target})~~~=~~~ 
{  {\rm ln}[1-P_{\rm Target}] \over  {\rm ln}[1-P_{\rm Success}({\cal T})] }~~~{\cal T}
\label{runtime}
\ee
is the mean physical run-time $\hat{\tau}_{\rm QA}$ for the successful ground-state searches
at a parametric given target success-rate $P_{\rm Target}$, which 
may be as close to unity as desired. 
The run times $\hat{\tau}_{\rm QA}$ naturally pend on 
the presence and properties of singularities 
at the quantum phase transition within the quantum 
partition function $Z_Q={\rm Tr} <e^{-\beta H_{QAC}}>$ 
as a function of $\lambda$ on $0 \le \lambda \le 1$. 
To a lesser extent one expects dependencies 
on the annealing schedule $\lambda (t)$ which we choose for
reasons of simplicity to be 
$\lambda (t)=t/{\cal T}$ i.e., the linear schedule.

\begin{figure}[htb*]
\centering{
a)\includegraphics[angle=-90,width=9.0cm]{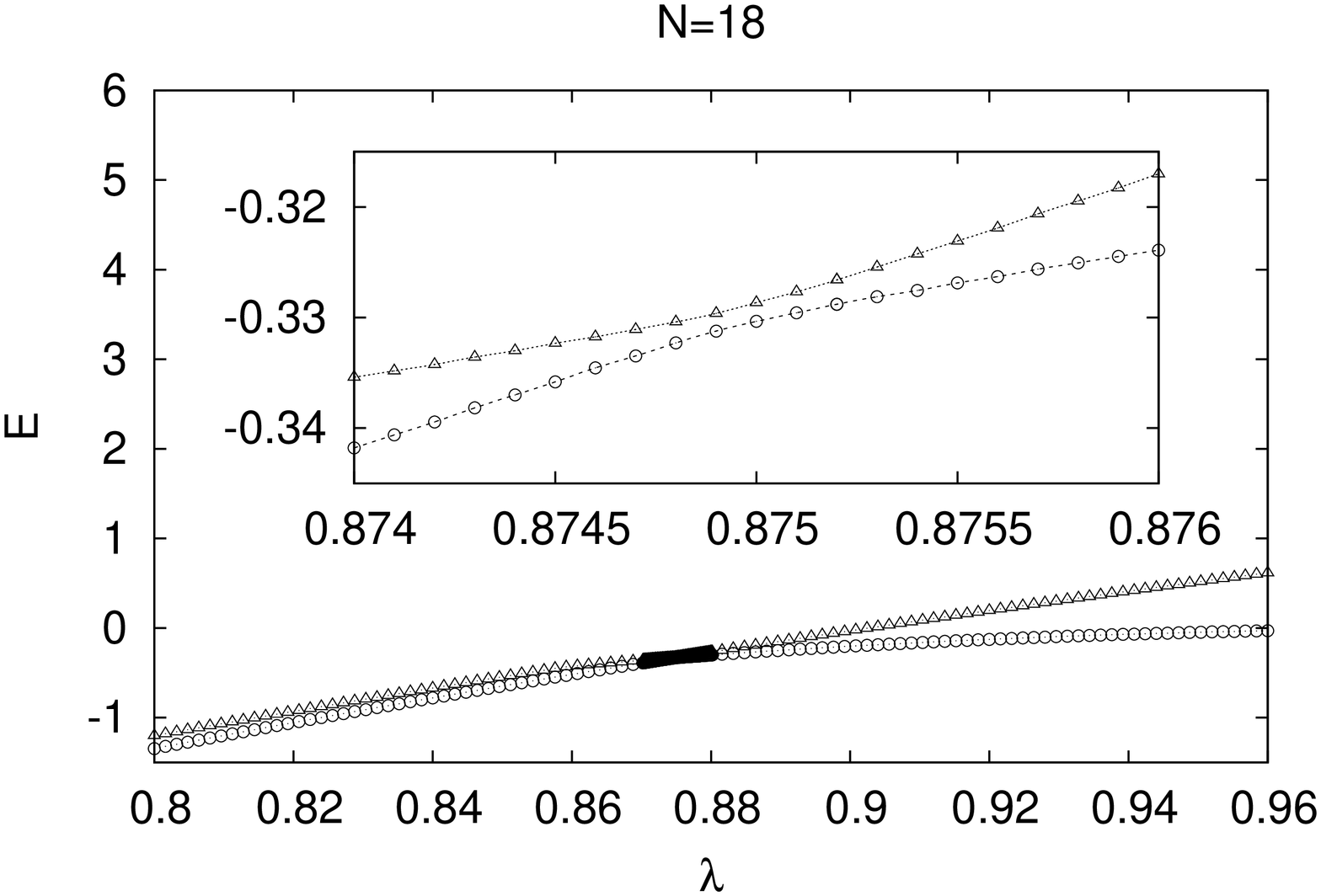} \\
b)\includegraphics[angle=-90,width=9.0cm]{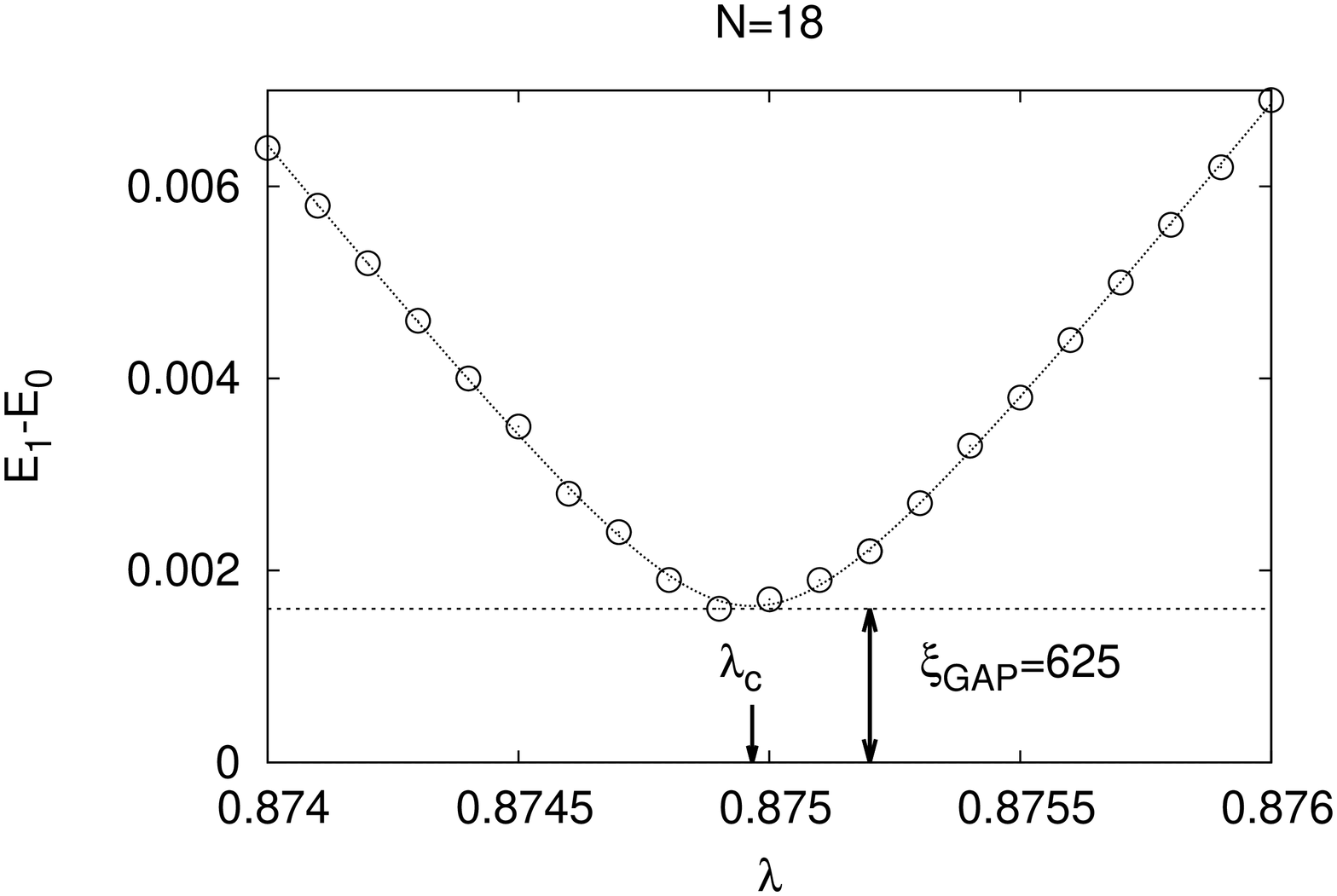}
\vspace*{0.0cm}
\caption{For a specific $N=18$ problem of median
complexity we display in a) the two lowest energy levels $E_0(\lambda)$
and $E_1(\lambda)$ as a function of $\lambda$. The inset in a)
displays the same data on a much smaller $\lambda$-interval. 
In addition the energy gap $\Delta E(\lambda)$ is displayed in b). 
The critical point $\lambda_c$ is marked by an arrow and 
the energy gap value corresponds to an horizontal line. The curve
represents a fit to the data with the form eq.(\ref{landau_zener_gap}).
The inverse energy gap parameter value $1/\Delta E_{\rm GAP}$
defines the gap-correlation length $\xi_{\rm GAP}:=1/\Delta E_{\rm GAP}$
at the value $\xi_{\rm GAP}=625$.}
}
\end{figure}

For realizations with a number of spins 
ranging in-between $N=3$ and $N=18$ we have 
computed levels of the instantaneous energy spectrum of the quantum adiabatic
Hamiltonian $H_{QAC}(\lambda)$ in eq.(\ref{HQAC}). We determine 
the two lowest energy levels:
the ground-state at energy $E_0$ and the first excited state at energy $E_1$.
We use the Lanczos algorithm and analyze either about $500$ or $1000$
problem realizations at each $N$ with $N \ge 10$ on a 
grid of discrete $\lambda$ values. 
For the smaller $N$ values we have less statistics at about $200$ problems
only. We consumed $34000$ core hours on a parallel computer.

For purposes of illustration and, for a representative 
median complexity instance at $N=18$,
we show in Fig. 2a) energy eigenvalues $E_0(\lambda)$ and  
 $E_1(\lambda)$, while Fig 2b) contains the energy gap $\Delta E=E_1-E_0$.
The limiting values of the gap in Fig 2b) are  
$\Delta E(\lambda=0)=2$ for the driver
Hamiltonian at $E_0(\lambda=0)=-N$ and $\Delta E(\lambda=1)=1$
at $E_0(\lambda=1)=0$ for the problem Hamiltonian and are outside
the $\lambda$ range of the figure. Limiting values have been checked for
any single problem. 
There is an important lesson to learn: 
The theory
exhibits just one quantum phase transition as signaled 
by a unique minimum energy gap value $\Delta E^i_{\rm GAP}$
at the transition point 
$\lambda_c^i$ for any given instance numbered by $i$ 
and, all the energy gap's
$\Delta E^i(\lambda)$ are in fact extremely 
well described by the three-parametric form
\be
\Delta E^i(\lambda)=\sqrt{ (~\Delta E^i_{\rm GAP}~)^2 +
(~[\lambda-\lambda_c^i]~\partial_\lambda\Delta E(\lambda_c)^i~)^2}
\label{landau_zener_gap}
\ee
in vicinity of the critical point
\footnote{The critical region at the LZ transition
is of size  $\delta \lambda \approx r 
\Delta E_{\rm GAP}/\mid \mid \partial_\lambda \Delta E \mid \mid$, where r is of ${\cal O}(10)$.
The fits are performed on the critical region, which requires fine 
$\lambda$-spacing on the $\lambda$-grid.} 
see e.g. the curve in Fig 2b).
This is the canonical form at an Landau-Zener (LZ) avoided level crossing. 
The presence of an isolated Landau-Zener level crossing at the 
quantum phase transition without
further structure on $0 \le \lambda \le 1$ 
relates to particular aspects of the spectrum which turn out to be simple: 
The model does not possess a possible cascade of avoided and actual
level crossings beyond $\lambda_c$ for $\lambda > \lambda_c$
as in \cite{tosatti_review_2006} and, also the energy 
gap does not close in vicinity of $\lambda=1$: The classical 
ground state is non-degenerate and as such there are no
tunneling events in-between degenerate vacua. The three parameters 
$\Delta E^i_{\rm GAP},\lambda_c^i$
and $\mid\mid \partial_\lambda\Delta E(\lambda_c)^i \mid\mid > 0$
are determined for any single instance $i$ of the problem set.

\begin{figure}[htb*]
\centering{
\includegraphics[angle=-90,width=8.5cm]{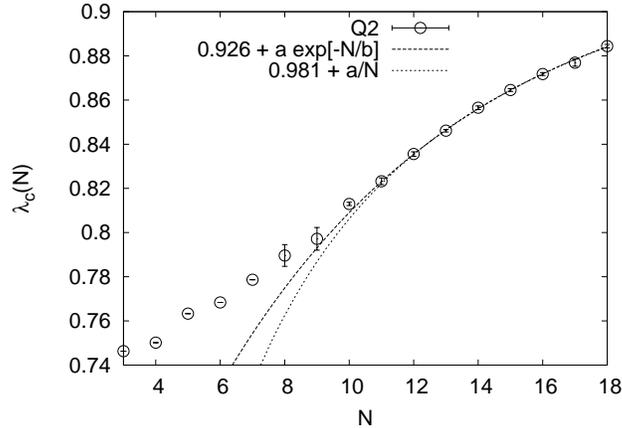}
\vspace*{0.0cm}
\caption{We display the quantum critical point $\lambda_c(N)$ in the
median $Q2$ as a function $N$, the circles in the figure. 
The two curves are explained in the text. They have the mathematical forms
as printed inside the figure.}
}
\label{fig_03}
\end{figure}

The positions of the avoided level crossings
$\lambda_c(N)$ have been determined in the median of the ensembles.
They are displayed in Fig. 3). However, and as there is no
established finite size scaling (FSS) theory for these class 
of phase transitions,
we feel unable to predict the critical point $\lambda_c$ 
in the thermodynamic limit! 
We note that we have experimented with the FSS forms 
$\lambda_c(N)=\lambda_c^{a)}+~a~{\rm exp}[-N/b]$ 
as well as $\lambda_c(N)=\lambda_c^{b)}+a/N$, parameterizing
exponential small as well as polynomial small corrections. 
Both Ans\"atze do in fact
yield reasonable $\chi^2_{dof}$-values if data with $N \ge 11$
are included into corresponding fits, see the 
two curves displayed in Fig. 3). 
However, the disparity in the parameters $\lambda_c^{a)}\simeq 0.93$
or $\lambda_c^{b)}\simeq 0.98$ reflects our
ignorance on the critical point position $\lambda_c$.

The obvious presence of a 
single Landau Zener avoided level crossing
at the critical point $\lambda_c$ 
leads us to consider Landau Zener theory  \cite{Landau_1932,Zener_1932}
for the dynamical success rates $P_{\rm Success}({\cal T})$
of eq.(\ref{runtime}), which 
becomes exact in the limit of infinite slow adiabatic state changes. 
For finite time schedules there will be corrections due to transitions
to higher energy eigen-states, which in a first step are dropped here.
Following these works the two energy level approximation 
to the diabatic success rate is 
\be
P_{\rm Diabatic}=e^{-\gamma}
\ee
with 
\be
\gamma= { 2 \pi \over {h} } 
{\Delta E_{\rm GAP}^2 \over \mid \mid  \partial_\lambda \Delta E(\lambda) \mid_{\lambda=\lambda_c}  {d \over dt} \lambda(t) \mid_{t=t(\lambda_c)} \mid \mid }.
\ee
Using the adiabatic success rate $P_{Success}=1-P_{\rm Diabatic}$ one then has
\be
\hat{\tau}_{\rm QA} (P_{\rm Target})~=~ {\rm ln}[{1\over1-P_{\rm Target}}]{\cal T \over \gamma}
\label{runtime2}
\ee
and inserting the linear schedule one finds
\be
\hat{\tau}_{\rm QA} (P_{\rm Target})~=~ {\rm ln}[{1\over1-P_{\rm Target}}]~({h \over 2\pi})~[{ \mid \mid \partial_\lambda \Delta E(\lambda_c) \mid \mid \over \Delta E_{\rm GAP}^2}].
\label{runtime3}
\ee
During this work we will drop all irrelevant factors from $\hat{\tau}_{\rm QA}$
and use
\be
\tau_{\rm QA}~=~{\mid\mid \partial_\lambda \Delta E(\lambda_c) \mid\mid \over \Delta E_{\rm GAP}^2}
\label{runtime4}
\ee
as the primary physical observable: the run-time 
which quantifies quantum search complexity.
It is understood that physical dimensions and trivial pre-factors
can be restored. The quantities $\tau^i_{\rm QA}$ are easily 
determined from the fits to the
energy gap see eq.(\ref{landau_zener_gap}) for 
any problem $i$ of the problem set. 
Finally we also
introduce the root 
\be
\Xi_{\rm LZ}~=~\sqrt{\tau_{\rm QA}}
\label{xilz}
\ee
and the static gap correlation length
\be
\xi_{\rm GAP}~=~{1 \over  \Delta E_{\rm GAP}}
\ee
which in the path integral formulation 
denotes the maximal correlation length in-between spins
at the quantum phase transition. 
For linear schedules and regular coefficients
$\mid\mid \partial_\lambda \Delta E(\lambda_c) \mid\mid$ 
the singular part in the run-time then
is carried by the singular contribution in $\xi_{\rm GAP}$
\be
{\rm ln} \tau_{QA}~=~2~{\rm ln}\Xi_{\rm LZ}~=~2~{\rm ln}\xi_{\rm GAP}+
{\rm regular~terms}.
\label{regular_terms}
\ee
The two hereby denotes the quadratic Landau Zener 
run-time pole, which dominates the run-time as the 
energy gap closes.

The run-times  $\tau^i_{\rm QA}$ and correlation length 
$\xi_{\rm GAP}^i$ as well as $\Xi_{\rm LZ}^i$ all are distributed
on the problem set as labeled by $i$.
Typically we can only afford twelve bins in histograms
at $m_{\rm Problem}=1000$ while otherwise 
the resulting PDF's turn out to be too irregular. 
We also apply statistical errors $\epsilon$ of 
magnitude $\epsilon=\sqrt(m)$ if a bin has $m$ entries.

We remark that the problem set generation as
given in  eq.(\ref{generating_function})
picks improbable problems at $\mu=1$ from 
a distribution that otherwise and at the satisfiability threshold 
of 2SAT is centered at large $\mu$-values and therefore catastrophic
statistics can be applicable. We consider here a two parametric 
class of Weibull $W_k$ and Frechet $F_k$ functions
\be
W_k(x)~=~{\cal  N}^{-1}~(x/x_0)^{(k-1)}~e^{-(x/x_0)^k}~
\label{Weibull}
\ee
\be
F_k(x)~=~{\cal  N}^{-1}~{(x_0/x)^{(k+1)}}~e^{-(x_0/x)^k},
\label{Frechet}
\ee
which pend on parameters $k$ and scales $x_0$.

\section{Simulated Annealing}

We have employed simulated annealing (SA) 
\cite{Kirkpatrick_Gelatt_Vecchi_1983}
as an additional and conceptually
different measure of complexity for the considered problem set. 
In its bare version simulated annealing is just an algorithm to 
solve problems. However, and as simulated annealing 
employs the canonical partition
function $Z_{\rm can}(\beta=T^{-1})$, where 
$T$ is the temperature, one can expect
that statistical properties determine the run-times in {\it simulated}
annealing.
Again there remains the fundamental issue in how far
Monte Carlo pseudo-dynamics approximates physical 
dynamics under the solution of the equations of motion.
For a failure under more physical dynamics than Monte Carlo namely
Fokker Planck dynamics see \cite{fokker_planck_02008}. 

In SA we set up the canonical partition function 
$Z_{\rm can}=\sum_{\rm conf}~{\rm exp}~[-\beta H_p]$
at inverse temperature $\beta=T^{-1}$ 
where $H_P$ is the problem Hamiltonian.
We choose a random initialized spin-configuration and perform
local Metropolis spin updates in a multi-spin coded computer 
program \cite{Wansleben_1984,Bhanot_1986}.  
We employ compute time farming on a parallel computer
with a parallel random number 
generator of 
Marsaglia \cite{Marsaglia_02003}. 
The annealing procedure is similar to the one used in 
 \cite{Neuhaus_on_hard_problems}, however we use a different schedule in 2SAT.
Each annealing trajectory is started at the high temperature
\be
T_{\rm 0}~=~10
\ee
and terminates after $100$ Sweeps i.e., $100 \times N$ Monte 
Carlo steps where $N$ is the spin number.
We use a multiplicative temperature schedule
\be
T_{\rm New}=~0.7847~T_{\rm Old}
\ee 
that lowers the temperature after each sweep, and after $20$ 
sweeps reaches low temperature $T=0.1$ already. 
We repeat the annealing trajectories $64000$ times
with different random numbers 
and determine the mean success probability 
$P_{\rm Success}^{SA}$ of successful 
ground-state searches after the sweep $100$.
Our measure of SA search run-time is  
\be
\tau_{\rm SA}~=~~{{\rm ln}[1-P_{\rm Target}^{SA}] 
\over {\rm ln}[1-P_{\rm Success}^{SA}]}~~\times~~100~~\times~~N~~[{\rm Monte~Carlo~Steps}],
\label{sa_run_time}
\ee
at target success rate one-half : $P_{\rm Target}^{}={1 \over 2}$.
The procedure is repeated for a possible $1000$ 
realizations and at all values of $N$.

\section{Success Probability Distributions}

From the point of view of a quantum 
annealing PDF's of 
gap correlation length values $PDF(\xi_{\rm GAP})$ 
are not direct measurements. Quantum annealers 
yield measurements for success-rates $P_{\rm Success}^i$ 
of instances at problem index $i$
and, if statistics over an ensemble is accumulated: the
probability distribution function $PDF(P_{\rm Success})$ is
a measurable observable. The latter probability contains 
a static measure of run-time complexity 
which then under the specific dynamics
is propagated into successes.

Within LZ-theory we can simply separate statics and 
dynamics, see the 
final result in eq.(\ref{fold}): At first we note that $\Xi_{\rm LZ}$
of eq.({\ref{xilz}}) is a {\it static} quantity which we will assume
to have a probability distribution
\be
{\rm PDF}(\Xi_{\rm LZ})~{\rm d}\Xi_{\rm LZ}.
\ee 
The function ${\rm PDF}(\Xi_{\rm LZ})$ can numerically 
be estimated here also by binning, and 
as the regular contributions of eq.(\ref{regular_terms}) 
turn out to only have smoothly varying dependencies,
its overall shape is close to the one of ${\rm PDF}(\xi_{\rm GAP})$.
At second we observe that
\be
P_{\rm Success}~=~1~-~e^{-\frac{R}{(\Xi_{\rm LZ})^{\bf 2}}}
\label{psuccess}
\ee 
with 
\be
R={2 \pi \over h {\cal T}}~~~{\rm:~LZ-theory}
\ee
constitutes a one parametric map from the static quantifier
to the success rate $P_{\rm Success}$ as a 
function of $R$. 
It is important to realize here that $R$ is a free parameter for
annealing devices that may be scaled by scale factors 
$s$ to any value $sR$ either by scaling 
the physical annealing time ${\cal T} \rightarrow s^{-1}{\cal T}$,
or by ''formally'' combining $s^{-1}$ repetitions of single annealing
runs into one combined success probability! A short calculation
yields for the inverse map 
\be
\Xi_{\rm LZ}~=~R^{{1 \over 2}}~\lbrace-{\rm ln}(1-P_{\rm Success})\rbrace^{-{1 \over 2 }}
\label{inverse}
\ee 
and for the Jacobian 
\be
\mid\mid \partial_{P_{\rm Success}}\Xi_{\rm LZ} \mid \mid~=~
{1 \over 2}~R^{-2}~\Xi_{\rm LZ}^3~e^{+\frac{R}{\Xi_{\rm LZ}^2}}.
\ee 
The desired probability distribution function 
${\rm PDF}(P_{\rm Success})~{\rm d}P_{\rm Success}$
can now be written down and up to an irrelevant normalization factor ${\cal N}^{-1}$
is 
\be
{\rm PDF}(P_{\rm Success})~{\rm d}P_{\rm Success}~=~ \break
{\cal N}^{-1} \times {\rm PDF}(\Xi_{\rm LZ})
\times \Xi_{\rm LZ}^3 \times e^{+\frac{R}{\Xi_{\rm LZ}^2}} \times {\rm d}P_{\rm Success}
\label{fold}
\ee
where it is understood that the inverse eq.(\ref{inverse}) $\Xi_{\rm LZ}=\Xi_{\rm LZ}(P_{\rm Success})$
is applied. This is an interesting finding and a comment is in order:

If from the functions of eq.(\ref{Weibull}) and eq.(\ref{Frechet})
one singles out the Frechet function $F_k$ at k=2
\be
{\rm PDF}(\Xi_{\rm LZ})~{\rm d}\Xi_{\rm LZ}~=~{\cal  N}^{-1}
~({\Xi_{\rm LZ}^0 \over \Xi_{\rm LZ}})^3~e^{-(\Xi_{\rm LZ}^0/\Xi_{\rm LZ})^2} 
~{\rm d}\Xi_{\rm LZ}
\ee
then terms $(\Xi_{\rm LZ})^{-3}$ cancel with corresponding terms from the Jacobian
and, one can find a value of $R=(\Xi_{\rm LZ}^0)^2$ 
such that ${\rm PDF}(P_{\rm Success})={\rm const}$ i.e.,
is constant. We find that quantum 
systems with with Frechet $F_{k=2}$ distributed 
${\rm PDF}(\Xi_{\rm LZ})$
and with an LZ quantum phase transition
ought to exhibit constant
success probability distributions in quantum annealing
for linear time schedules, if the annealing times ${\cal T}$
are tuned to the point of mean success $<P_{\rm Success}>={1 \over 2}$. 
Scanning the parameter space $k$ of $W_k$ eq.(\ref{Weibull}) 
and $F_k$ eq.(\ref{Frechet}) in-between
$k=1$ and $k=3$ as models for 
${\rm PDF}(\Xi_{\rm LZ})~{\rm d}\Xi_{\rm LZ}$, and always 
tuning $R$ to the point of mean success one half, we find for the Frechet case
bimodal success probability distribution functions 
at $k>2$ while for $k<2$ these
turn out to be uni-modal. A corresponding watershed line does not exist for the
Weibull case for which all probability distributions are bimodal in $1<k<3$.
Within the scope of this work we will numerically determine
values $\Xi_{\rm LZ}$ and PDF's for the given problem 
set and with the help of eq.(\ref{fold}) fold them into
corresponding success probability distributions. 



\section{Findings}

\subsection{Problem Set Correlations of Observables}

\begin{figure}[htb*]
\centering{
\includegraphics[angle=-90,width=10.0cm]{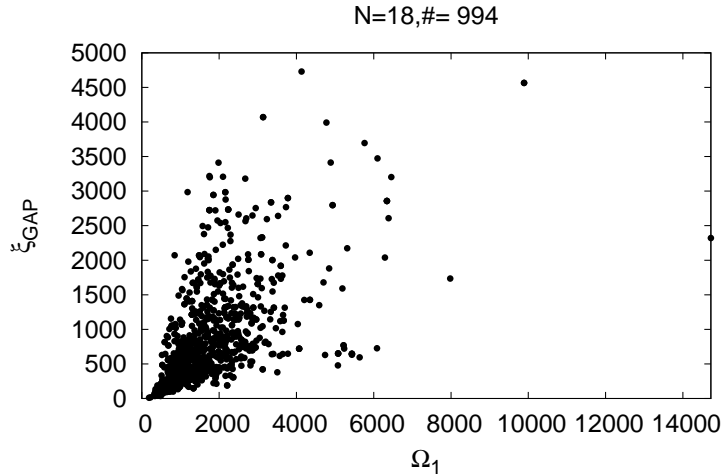}
\vspace*{0.0cm}
\caption{For the $N=18$ problem set we display for each problem
the density of states $\Omega_1$ at $E=1$  
and the quantum gap correlation length $\xi_{\rm GAP}$. There
are $994$ data points in the figure. The data appear to be to a large extent 
un-correlated.}
}
\end{figure}

Our numerical study provides a wealth
of observables that quantify quantum 
as well as classical search 
complexity for each incidence 
on the problem set. These are for QA the
gap correlation length $\xi_{\rm GAP}$, the
run-time $\tau_{QA}$, its root $\Xi_{\rm LZ}$ and success rates
$P_{\rm Success}$. For SA run-times in Sweeps $\tau_{SA}$, 
corresponding success-rates $P_{\rm Success}^{\rm SA}$
and a free energy $\Omega_1=\Omega(E=1)$ are given. 
The observables either are exact as in the case of $\Omega_1$,
carry numerical errors of relative magnitude 
$\delta o/ {\rm o} \approx {\rm O}(10^{-4})$ like for $\xi_{\rm GAP}$, 
or have small statistical errors as for $\tau_{\rm SA}$.
We transform any set of success probabilities 
$P_{\rm Success}^i$ via 
\be
P_{\rm Success}^i \rightarrow 1-(1-P_{\rm Success}^i)^R 
\ee
to the point of mean equal success rate one half 
$<P_{\rm Success}>={1 \over 2}$
by an appropriate choice of $R$. The considerations 
on the shapes of success PDF's of the last section then apply. 
One of the interesting issues now are correlations in-between 
observables for a given problem set at fixed $N$, which we choose 
to discuss on a semi-quantitative level.

\begin{figure}[htb*]
\vspace*{0.0cm}
\centering{
\hspace*{0.0cm}
\includegraphics[angle=-90,width=10.5cm]{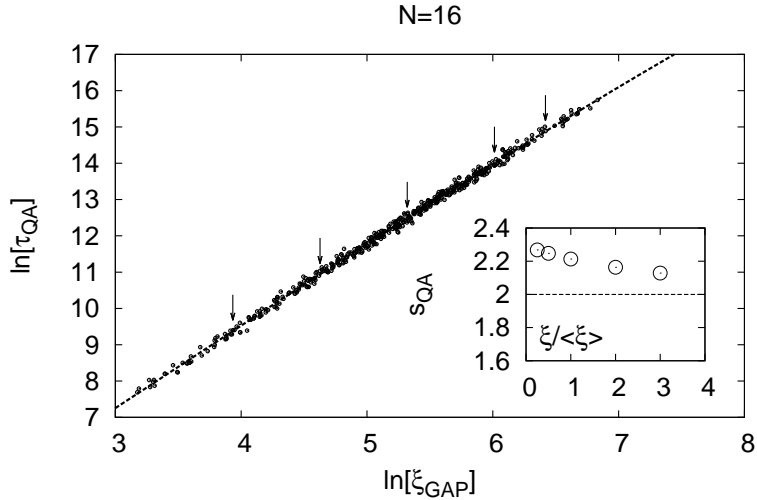}
}
\vspace*{1.0cm}
\caption{Scatter plot of correlated and logarithmic
$\tau_{\rm QA}$ and  $\xi_{\rm GAP}$ 
observables for the problem set at $N=16$.
The curve interpolating the data corresponds to a 3-rd order
polynomial fit. 
The inset of the figure displays a numerical estimate of the 
derivative, which is evaluated
at the arrow positions and approaches the LZ value two for hard problems.
The x-axis in the inset is the correlation length in units 
of the median.}
\end{figure}

The run time behavior in SA and QA on
the problem set can, problem by problem
either be correlated, or 
correlations may fall apart 
in which case different run-time 
scaling in $N$ is possible.
Recalling $\xi_{\rm GAP}$ 
and $\Omega_1$  to be static 
quantifiers of QA and SA search complexity 
we display in Fig. 4) a scatter plot of 
both for $N=18$ spins. Similar figures
can be obtained at smaller $N$.
It is quite evident: The gap correlation length and the 
energy one degeneracy are to a large extent un-correlated.
For reasons of completeness we cite
the maximum gap correlation length
from the union of all problems, which 
at $N=18$ was found to have the value
\be
\xi_{\rm GAP}\mid_{\rm Maximal~Value~over~All} ~=~ {4730}~~~.
\ee
We remark that such large correlation length 
values constitute an impasse
to any Quantum Monte Carlo (QMC) simulation for the spectrum. 
Whether the same holds true for {\it simulated} quantum 
annealing studies with 
QMC methods, is an open question.

\begin{figure}[htb*]
\centering{
\includegraphics[angle=-90,width=10.5cm]{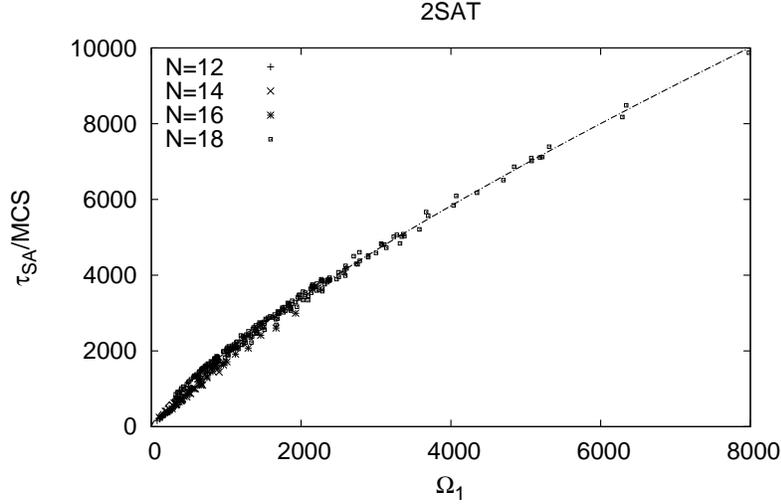}
\vspace*{0.0cm}
}
\caption{Scatter plot of correlated 
$\tau_{\rm SA}$ and  $\Omega_1$ 
observables for the problem set at $N=12,14,16$ and $N=18$. 
The interpolating curve in the figure is explained in the text.}
\label{fig_of_sa_run_time}
\end{figure}

The situation however, as far as correlations are concerned 
turns around from largely
un-correlated to {\it correlated} when correlations of 
static search quantifiers 
$\Omega_1$ respectively $\xi_{\rm GAP}$ 
with their corresponding dynamic counterparts, namely run-times 
$\tau_{\rm SA}$ and $\tau_{\rm QA}$ are considered.
On the quantum annealing side we display in  Fig. 5) 
$\tau_{\rm QA}$ and  $\xi_{\rm GAP}$ 
observables for the $N=16$ problem set 
in a common scatter plot at logarithmic scale. 
Both quantities show an almost linear
correlation in-between the logarithms, the small open circles in the figure.
A closer inspection yields, see the inset in Fig. 5), 
that the slopes $s_{\rm QA}$
of the logarithmic run-time derivatives  
$s_{\rm QA}=\partial_{{\rm ln}\xi_{\rm GAP}} {\rm ln}\tau_{\rm QA}$
slightly overshoot the quadratic LZ pole value: $s_{\rm QA}=2$. The
margin is about $15$ percent for easy problems 
at small $\xi_{\rm GAP}$. The difference to the pole value $s_{\rm QA}=2$
then vanishes for the very hard problems.
For the linear run-time schedules as considered here, these run-time corrections 
are induced by the coefficients $\mid\mid \partial_\lambda \Delta E(\lambda_c) \mid\mid$
of eq.(\ref{runtime4}). 
In the logarithmic derivatives they turn out to be regular.
We calculated the derivatives numerically. For this purpose
a 3-rd order polynomial is fitted to the data, which then is differentiated.

On the classical annealing side we display in  Fig. 6) $\tau_{\rm SA}$ and  
$\Omega_1$ observables for the $N=12,14,16$ and $N=18$ problem sets 
in a common scatter plot at linear scale. 
The data show a band of allowed
correlations, which is significant and is not caused by the statistical errors
of the SA simulation as errors are small. 
We suspect internal 
structure on the energy one surface relative to the ground-state in terms of 
spin flip distances and their stochastic distribution. 
We fit the correlation by the 
Ansatz $\tau_{\rm SA}=A\Omega_1^{1 + \Delta \alpha_{\rm SA}}$
and obtain $ \Delta \alpha_{\rm SA}=-0.16(1)$. A representative scaling form 
is displayed in Fig. 6), the curve in the figure. 
We mention, that the performance of different classical annealer's in case of Dwave
is used to question the quantum nature of the search process 
\cite{Browne_02014,Smolin_02013,Smolin_02014}.  
It is inconceivable that for our toy model classical
annealer's would not have $\Omega_1$ as static quantifier. 
If however classical dynamics can model Dwave, then 
criticism to the effect that the machine is operating classical 
has to be taken more serious. 

\begin{figure}[htb*]
\centering{
\includegraphics[angle=-90,width=7.5cm]{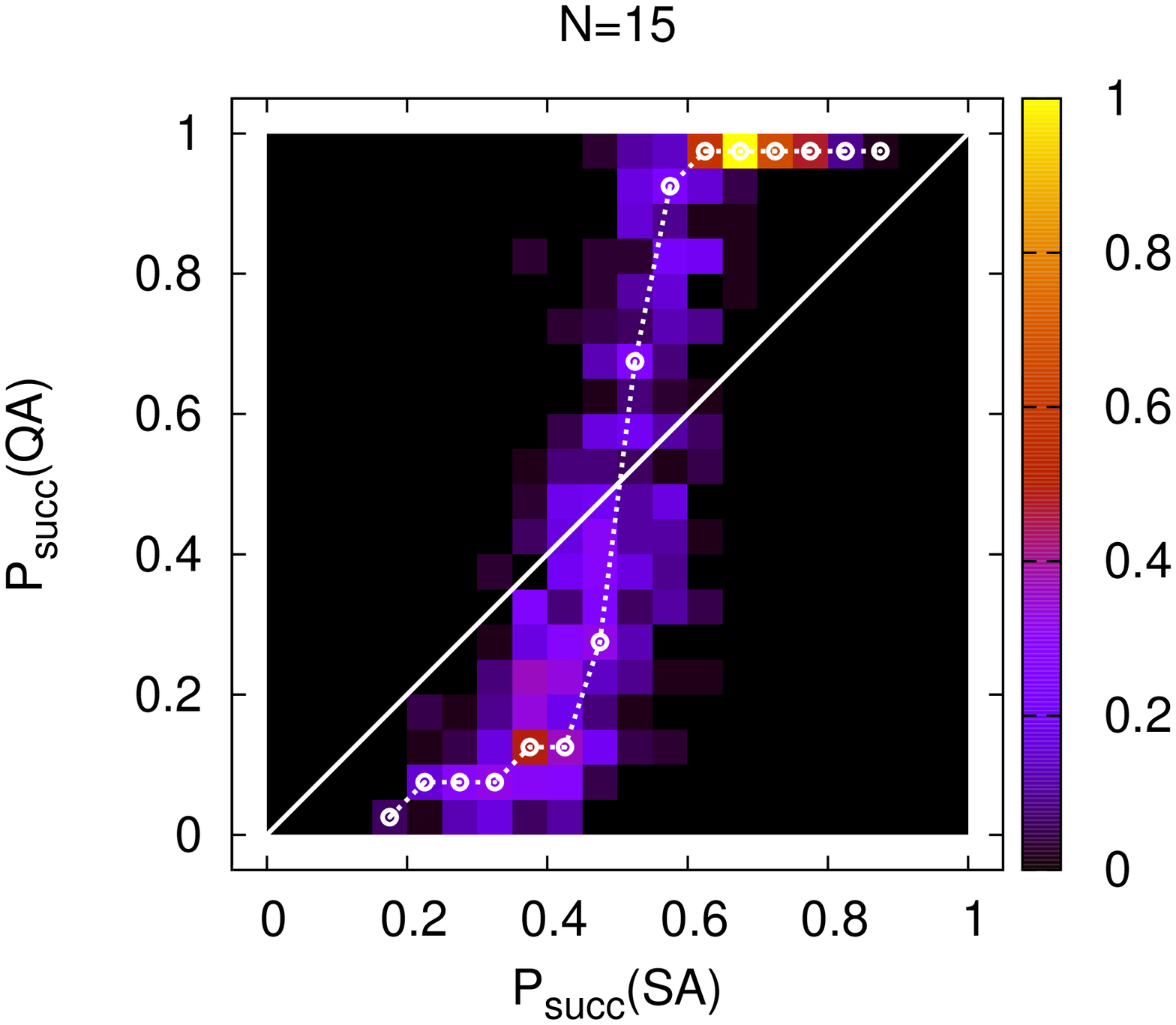}
\includegraphics[angle=-90,width=7.5cm]{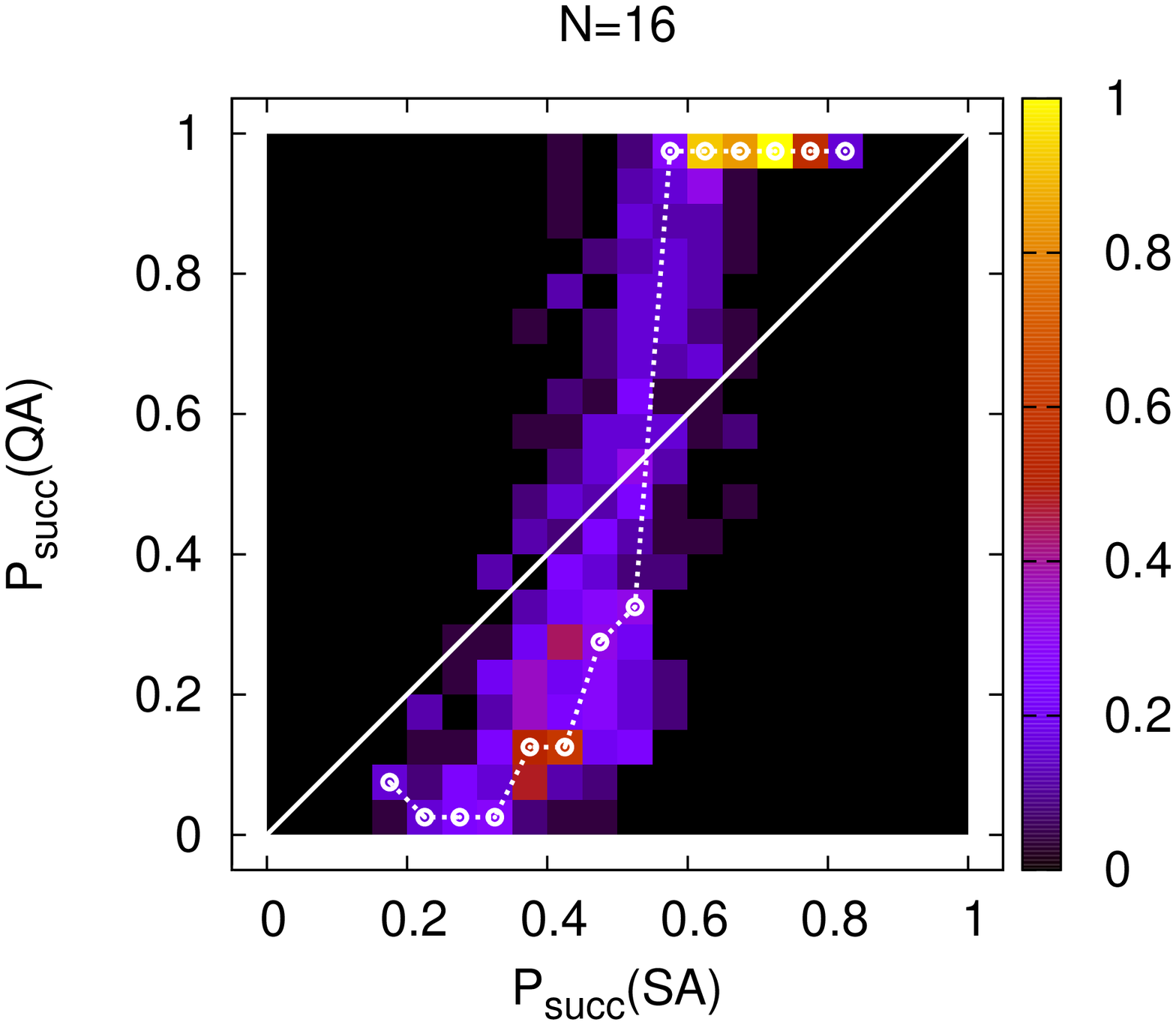} \\ 
\includegraphics[angle=-90,width=7.5cm]{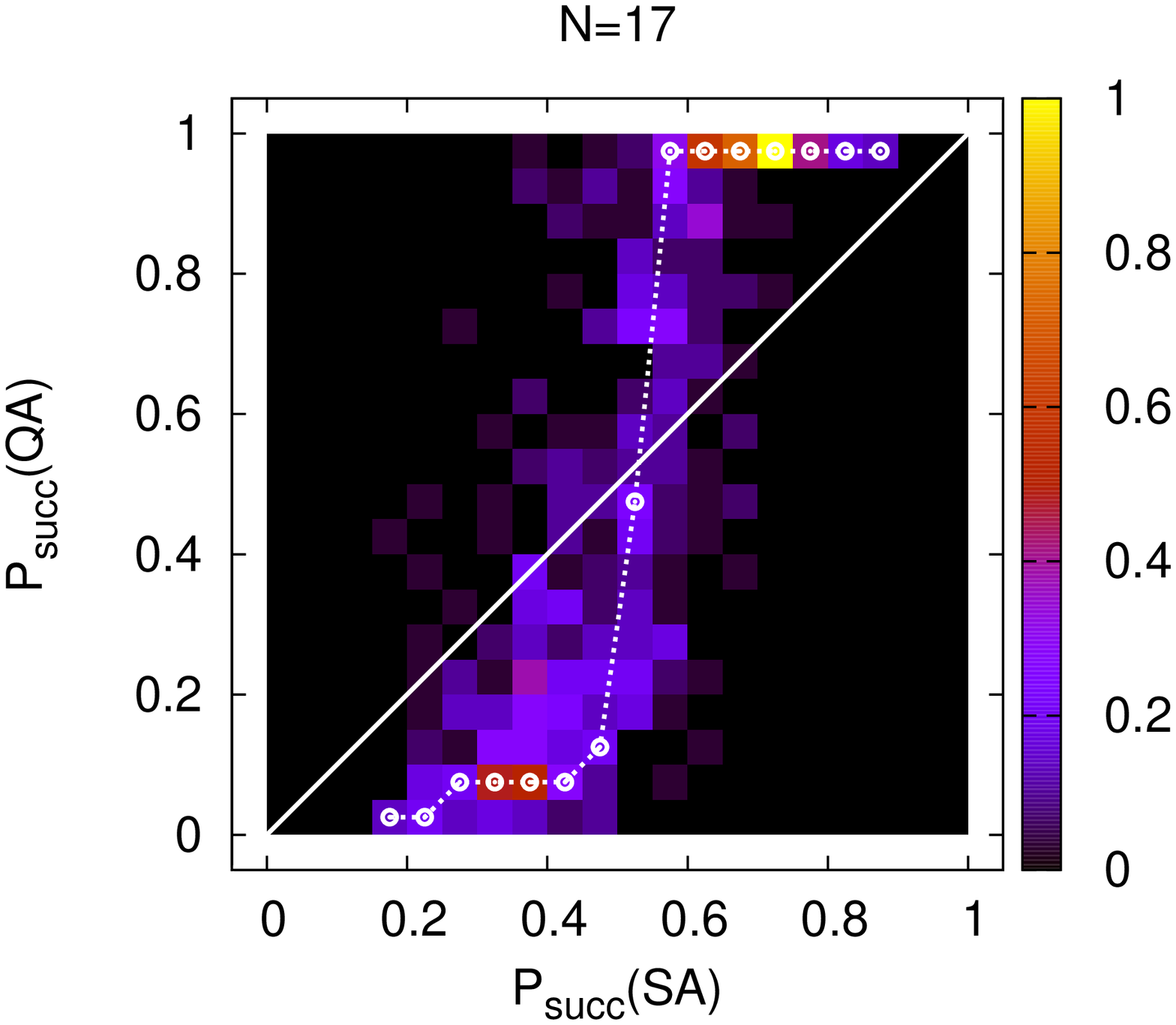}
\includegraphics[angle=-90,width=7.5cm]{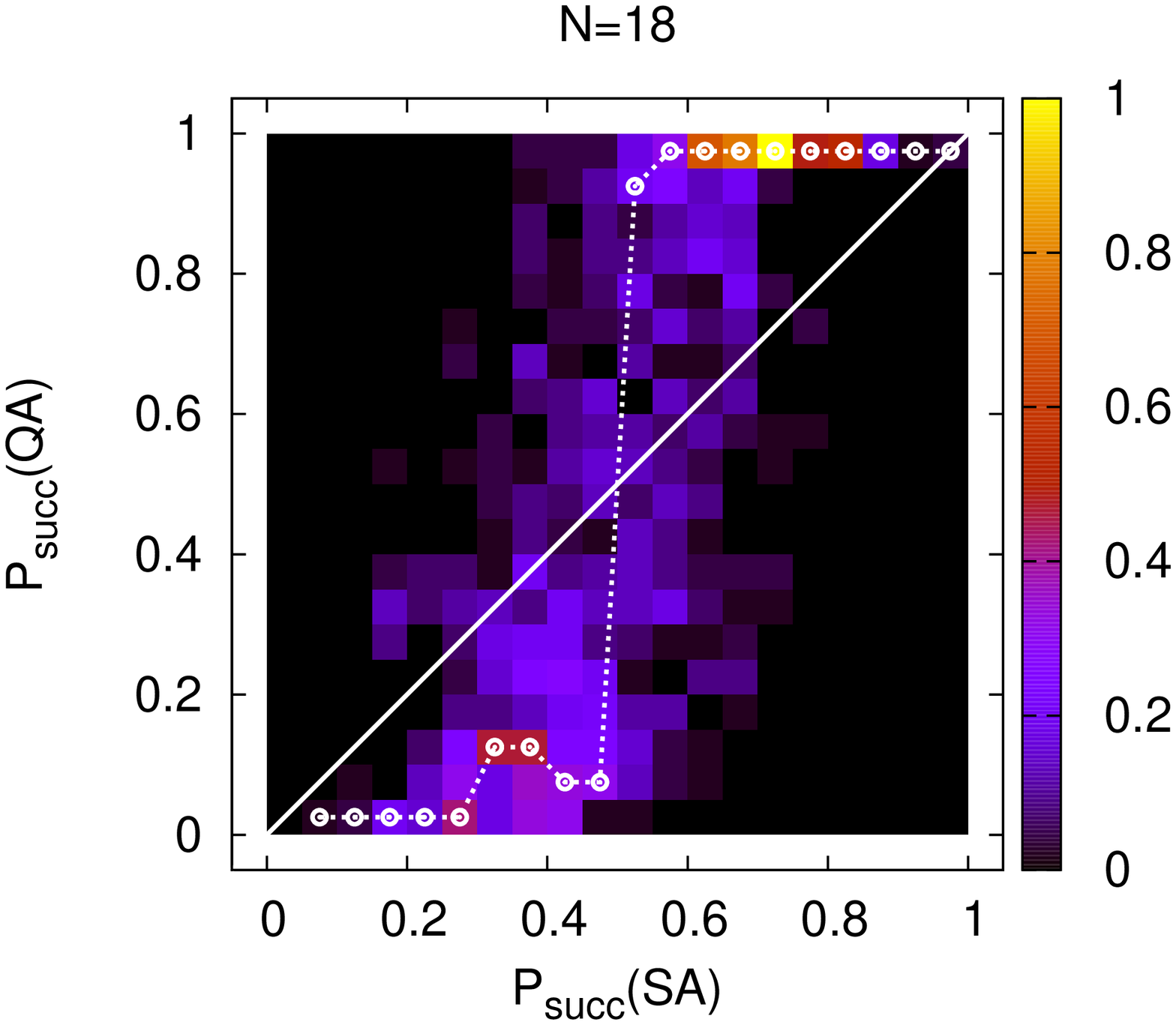}  
\vspace*{0.0cm}
\caption{Double histograms of correlated success 
rates $P_{\rm Success}^{SA}$ (SA) (x-axis) and $P_{\rm Success}$ (QA) 
(y-axis). Clock-wise and starting in the left upper corner
we present data for $N=15,16,18$ and $N=17$ problem sets.
The histograms have two local maxima at approximate positions
$(P^{\rm SA},P^{\rm QA}) \approx (0.7,1.0)$ and  $\approx (0.4,0.1)$.
Polygons in the figures indicate peak positions. The straight lines
in the figures mark the position of linear correlations.}
}
\end{figure}

\begin{figure}[htb*]
\centering{
\includegraphics[angle=-90,width=7.5cm]{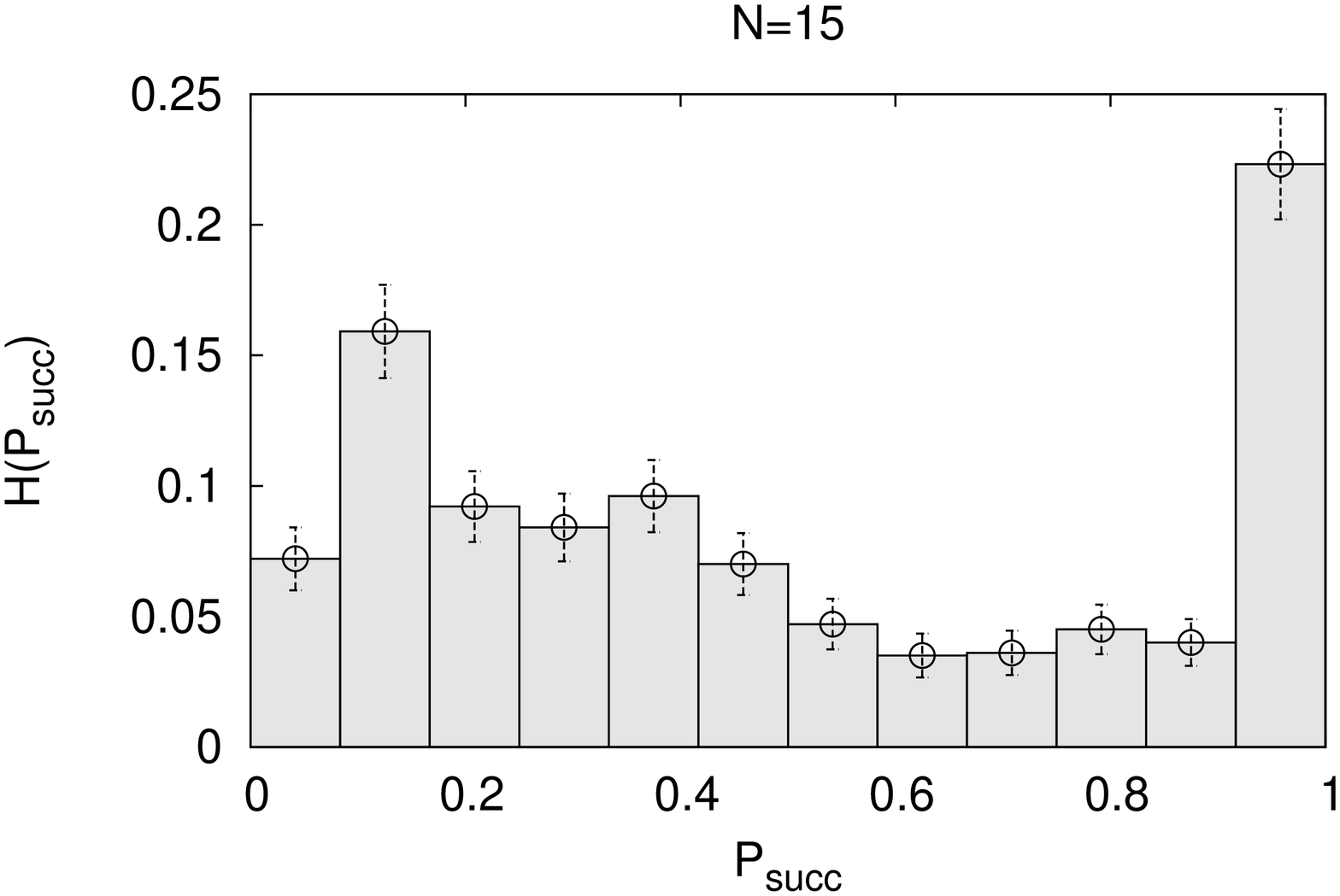}
\includegraphics[angle=-90,width=7.5cm]{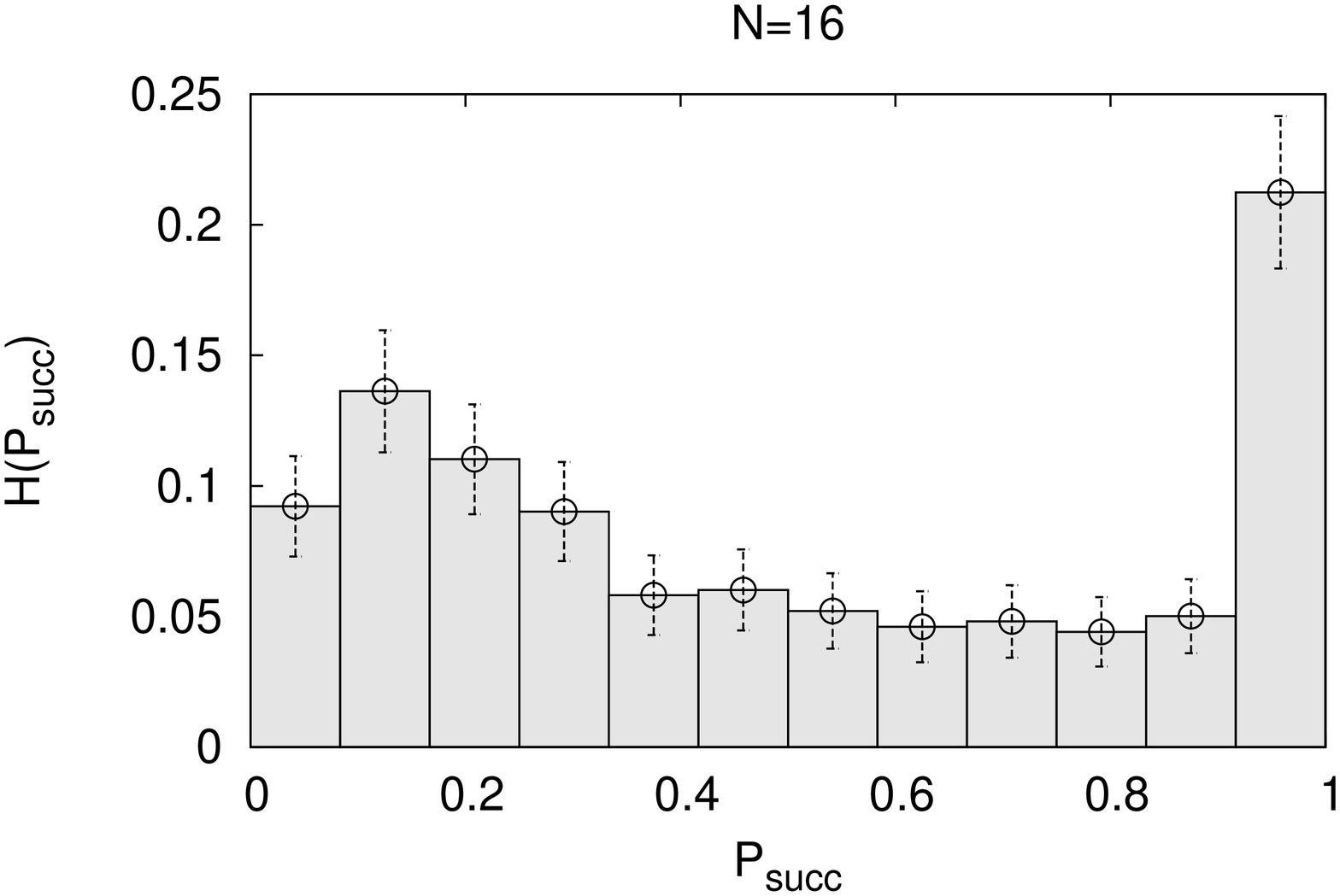} \\
\includegraphics[angle=-90,width=7.5cm]{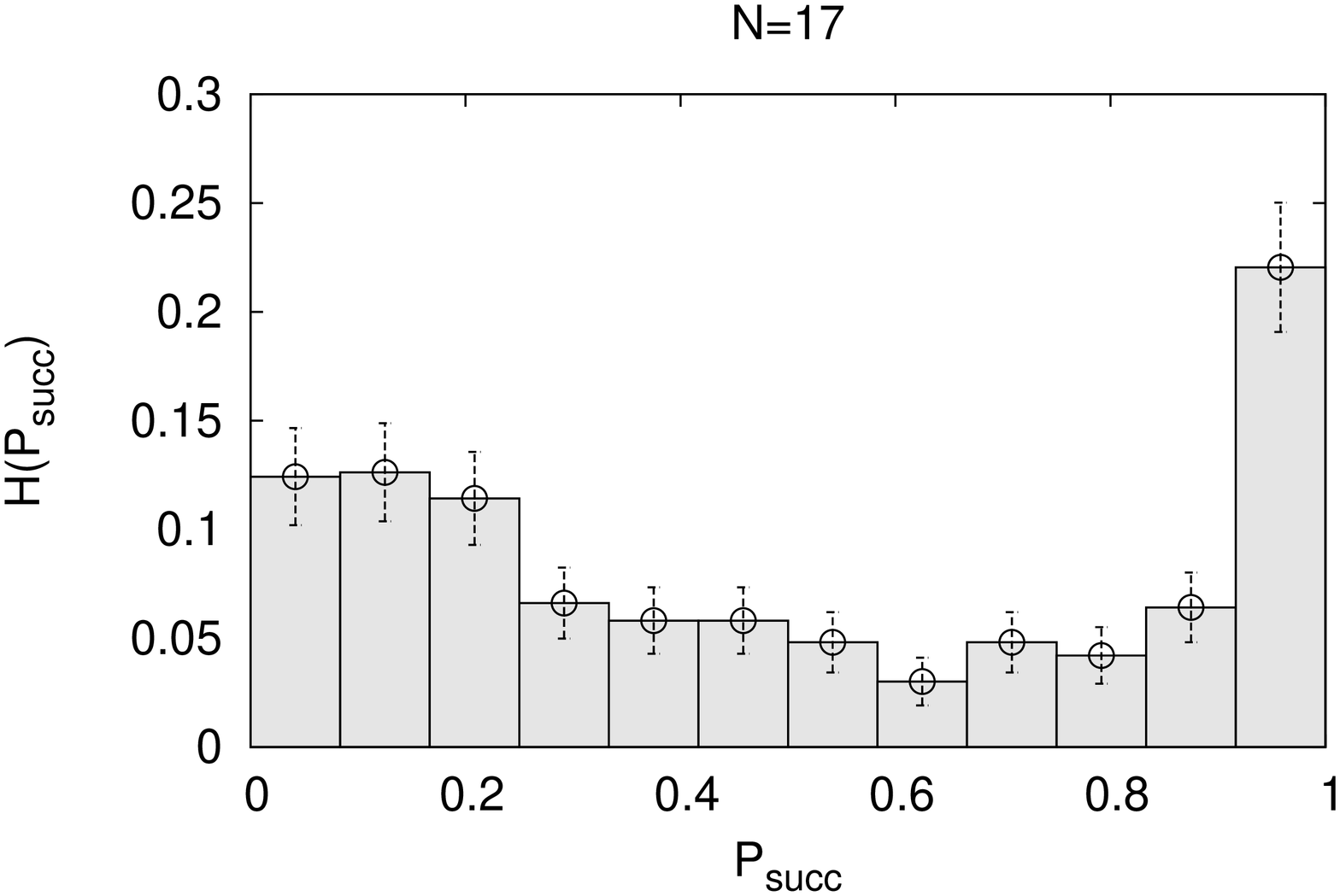}
\includegraphics[angle=-90,width=7.5cm]{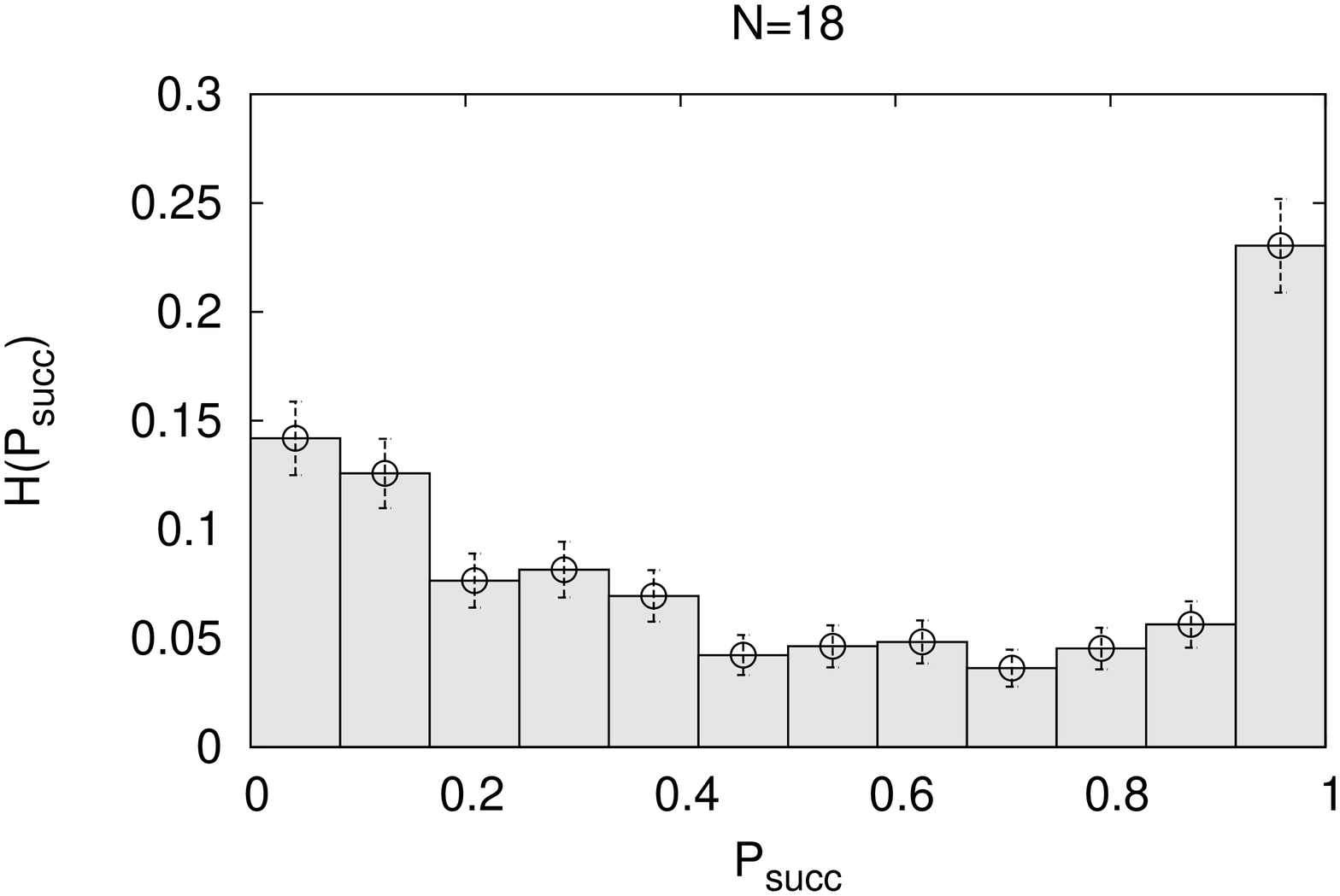}
\vspace*{0.0cm}
\caption{Success probability distributions $PDF(P_{\rm Success})$ for quantum annealing
on the problem set at $N=15,16,18$ and $N=17$ (clock-wise).}
}
\label{success_pdfs}
\end{figure}

Finally a comprehensive pictorial representation 
of different quantum versus classical
search complexities is obtained in terms of correlations in-between
success probabilities $P_{\rm Success}$ for QA
and $P_{\rm Success}^{\rm SA}$ for SA, which both are 
compactified observables on the 
interval $0 \le P \le 1$.
It is straightforward to construct a double histogram 
$H(P_{\rm Success}^{\rm SA},P_{\rm Success})$ on the problem set,
which counts entry's into bins and is normalized to unity 
at the bin of maximal 
frequency. The data are displayed in Fig. 7) for the largest spin numbers.
There is a band of events which parts form the linear 
correlation, see the straight
lines in Fig. 7), and clearly forms a step-like structure
at an value of $P_{\rm Success}^{\rm SA}\approx 0.5$. It is clear, and 
with reference to the arguments of the last section, that we can
attribute this step to the presence of catastrophic statistics  
in the distribution of the quantifier $\Xi_{\rm LZ}$ 
in eq.(\ref{fold}).
This issue and the 
probability distributions of  $\Xi_{\rm LZ}$, $\Xi_{\rm GAP}$ and of 
$P_{\rm Success}$ observables will be discussed in the next section.
The work \cite{Troyer_02013} presents similar success rate
correlations in-between Dwave and SA again. The step 
there, see Fig. 2c) of the paper, turns out to be less symmetric with a larger
amount of statistical noise. 
However, the theories definitely are different and nothing 
quantitative can be concluded from the comparison.

\subsection{Probability Distribution Functions, PDF's}

\begin{figure}[htb*]
\centering{
\includegraphics[angle=-90,width=7.5cm]{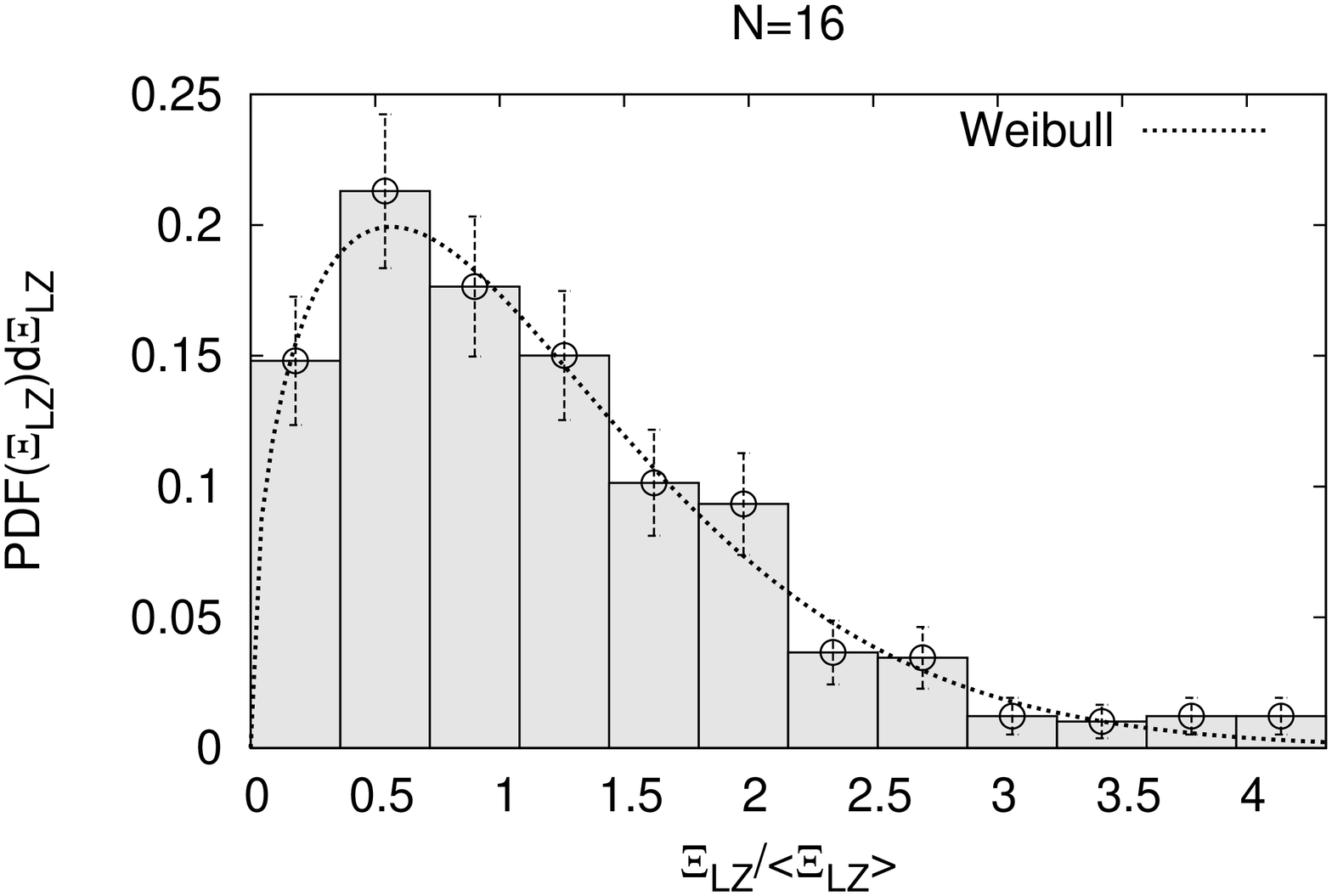}
\includegraphics[angle=-90,width=7.5cm]{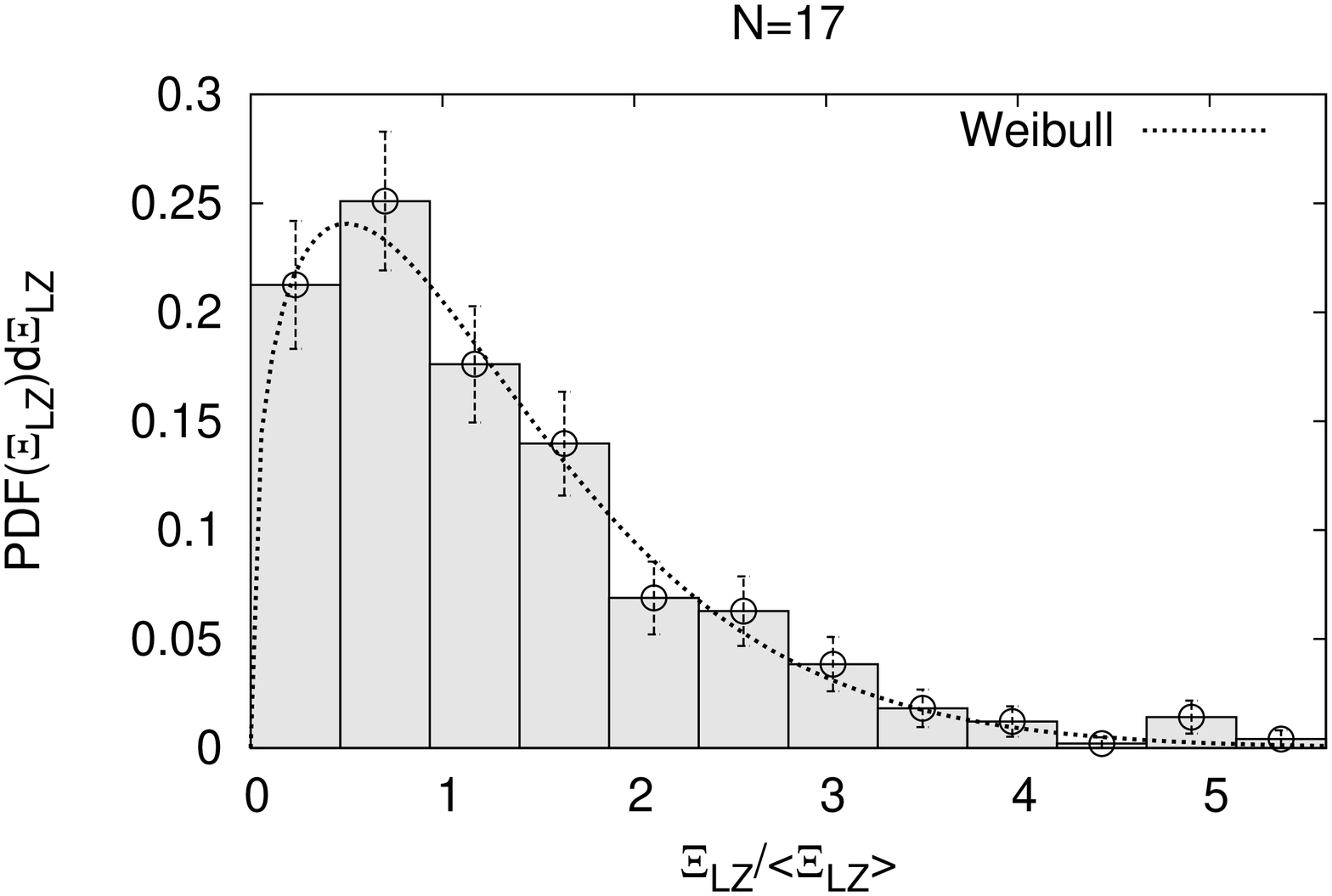} \\
\includegraphics[angle=-90,width=7.5cm]{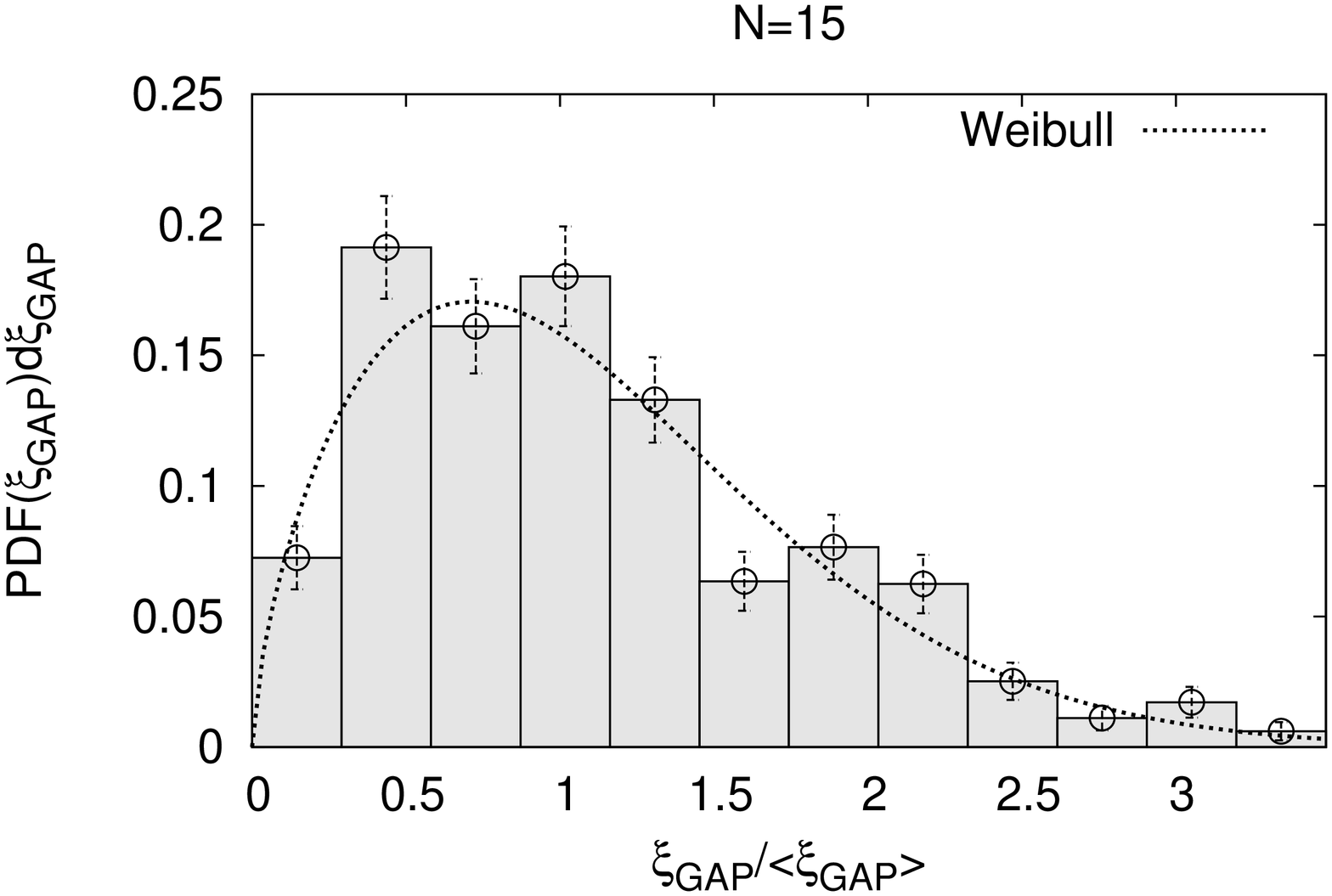}
\includegraphics[angle=-90,width=7.5cm]{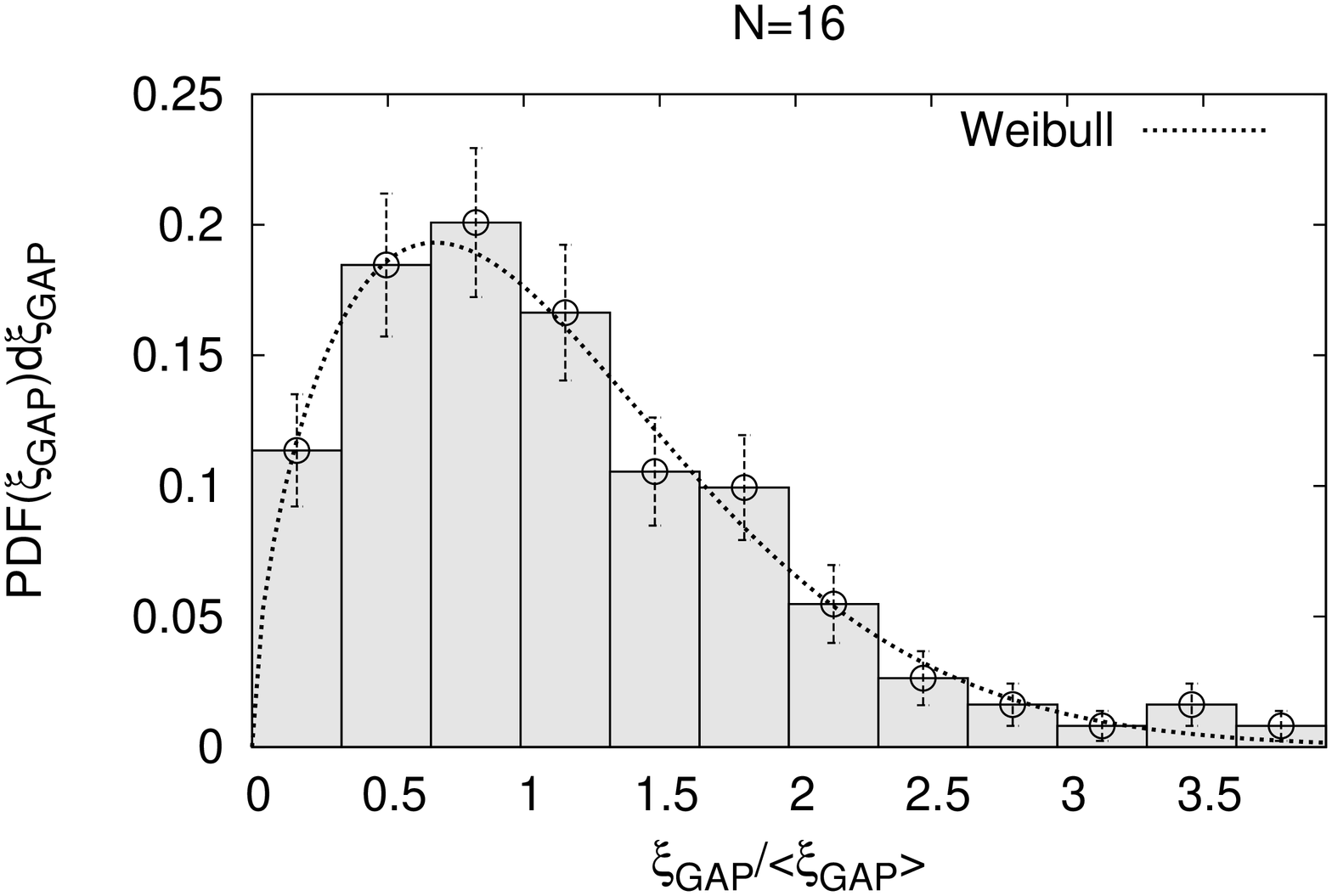} 
\vspace*{0.0cm}
\caption{PDF's for $\Xi_{\rm LZ}$, the square root of the run-time
$\Xi_{\rm LZ}=\sqrt{\tau_{\rm QA}}$, in the top row for $N=16$ and $N=17$.
PDF's for $\xi_{\rm GAP}$, the gap correlation length, in the bottom row for $N=15$ and $N=16$.
The symbol $<>$ denotes the median expectation value.}
}
\label{fig_09}
\end{figure}
~~~~~~~~~~~~~~~~~~~~~~~~~~~~~~~~~~~~~~~~~
\begin{table}[htb*]
\centering{
\begin{tabular}{|c|c|c|c|c|c|c|} 
\hline      
$N$ & $\chi^2_{dof}$ & $\xi_0/<\xi_{\rm GAP}>$ & $k_{\rm Weibull}$ & $\chi^2_{dof}$ & $\Xi_0/<\Xi_{\rm LZ}>$ & $k_{\rm Weibull}$ \\
\hline\hline
10 & 4.12 & 1.11(3) & 3.02(29) & 6.94 & 1.23(6) & 2.82(28)\\
11 & 3.88 & 1.18(4) & 2.84(20) & 2.28 & 1.24(4) & 2.19(14)\\
12 & 6.22 & 1.18(6) & 2.70(21) & 5.21 & 1.24(6) & 2.17(16)\\
13 & 3.53 & 1.26(5) & 2.31(14) & 3.00 & 1.27(5) & 1.90(11)\\
14 & 1.97 & 1.26(4) & 1.98(09) & 0.52 & 1.28(2) & 1.67(04)\\
15 & 2.44 & 1.25(6) & 1.65(09) & 1.36 & 1.26(5) & 1.40(07)\\
16 & 0.53 & 1.25(3) & 1.58(06) & 0.62 & 1.28(4) & 1.43(06)\\
17 & 0.36 & 1.34(3) & 1.40(05) & 0.71 & 1.36(5) & 1.34(06)\\
18 & 1.61 & 1.37(7) & 1.24(07) & 0.66 & 1.39(5) & 1.15(04)\\
\hline\hline
\end{tabular}
\caption{Fit results to probability distributions $PDF(\xi_{\rm GAP})$ 
in row $2,3,4$ and to probability 
distributions $PDF(\Xi_{\rm LZ})$ in row $5,6,7$.
The spin number $N$ is contained in row $1$. Each block of fits
contains $\chi^2_{d.o.f}$ values, scales and most important 
the Weibull parameter $k$ in row $4$ and row $7$.} 
}
\end{table}

The discretized numerical success probability 
distribution functions  $PDF(P_{\rm Success})$ 
of the quantum annealing processes on the problem set are
bimodal, they span the whole compact interval $[0 \le P \le 1]$
without any tendency of self-averaging at increasing $N$,
and are displayed in Fig. 8) for $N=15,16,17$ and $N=18$.
These distributions constitute theoretical predictions for quantum
annealer's that run the linear schedule in the large annealing time ${\cal T}$
limit. The success probability
distributions of simulated annealing (SA) are unimodal, as in the 
case of Dwave \cite{Troyer_02013,Troyer_02014}.

For the linear time schedule the static distributions of 
$\Xi_{\rm LZ}=\sqrt{\tau_{\rm QA}}$ are relevant, while the
$\xi_{\rm GAP}$ gap correlation length distributions 
at the quantum phase transition
anyhow are fundamental. Both kind of probability distributions are displayed
Fig. 9): the top row contains $PDF(\Xi_{\rm LZ})$ at $N=16$ and $N=17$, 
while the bottom row contains $PDF(\xi_{\rm GAP})$ at $N=15$ and $N=16$.
On the main diagonal of the figure at $N=16$ one can compare
$\Xi_{\rm LZ}$ and $\xi_{\rm GAP}$ PDF's for an 
identical problem set.
All distributions are plotted as a function of their corresponding
variable in units of the median as denoted by $<>$. One notices 
that the distributions of  $PDF(\Xi_{\rm LZ})$ and  $PDF(\xi_{\rm GAP})$ at $N=16$ 
do not differ much, when plotted as a function of median normalized x-coordinates.
We have performed
fits to the distribution data employing the catastrophic 
functions $W_k$ and $F_k$ of eq.(\ref{Weibull}) and eq.(\ref{Frechet}).
The qualities of the fits single out Weibull functions $W_k$
as the better models for data and, thus we fit e.g. the Ansatz
\be
PDF(\xi_{\rm GAP})~=~A_0~(\xi_{\rm GAP}/\xi_0)^{(k-1)}~e^{-(\xi_{\rm GAP}/\xi_0)^k}~~~~[{\rm Weibull}]
\ee
to the $\xi_{\rm GAP}$ distribution. The fit-results for 
parameters $k[\xi_{\rm GAP}]$, scales $\xi_0$ 
and correspondingly $k[\Xi_{\rm LZ}]$ and scales $\Xi_0$ are all 
collected in Table 2), 
together with the $\chi^2_{d.o.f}$ values for the fits. The 
fitted curves are also displayed
in Fig. 9), see the curves in the figure. Most importantly we
find that for our largest systems 
at $N=14,...,18$ Weibull functions 
describe the PDF's
to good precision, see the $\chi^2_{d.o.f}$ 
values in Table 2). This implies thin catastrophic distribution function 
tails in the variables $\Xi_{\rm LZ}$ or $\xi_{\rm GAP}$.
The value of $k$ slowly decreases from $k=1.65$ respectively 
$k=1.40$ at $N=15$, to $k=1.24$ respectively 
$k=1.15$ at $N=18$ and possible assumes $k=1$ at infinite $N$. 
None of the distributions has reached its stationary shape.
We mention that a duality in-between 
$W_k$ and $F_k$ implies, that 
Weibull distributions for gap correlation length values 
turn into Frechet distributed {\it energy gap} distributions 
at $k$. Neither the semi Poisson 
distribution of nearest neighbor levels
in randomly distributed energies
$P(\Delta E)d\Delta E \propto \Delta E{\rm exp}[-\Delta E/\Delta E_0]$
, nor Wigner's surmise  \cite{Wigners_surmise}
$P(\Delta E)d\Delta E \propto \Delta E{\rm exp}[-(\Delta E/\Delta E_0)^2]$
of energy gaps aka random matrix theory for quantum level
repulsion in the bulk are appropriate models here. 
If $k$ actually takes the asymptotic value $k=1$, then 
$P(\Delta E)d\Delta E \propto (\Delta E)^{-2}{\rm exp}[-\Delta E_0/\Delta E]$
is predicted with a fat tail in $\Delta E$ ! Znidaric in \cite{Znidaric_2005_eigenvalue_statistics}
found a Poisson distribution for the energy gaps in 3SAT with a thin tail in  $\Delta E$ however.
 
\subsection{Run Time Singularities}

\begin{figure}[htb*]
\centering{
\includegraphics[angle=-90,width=8.5cm]{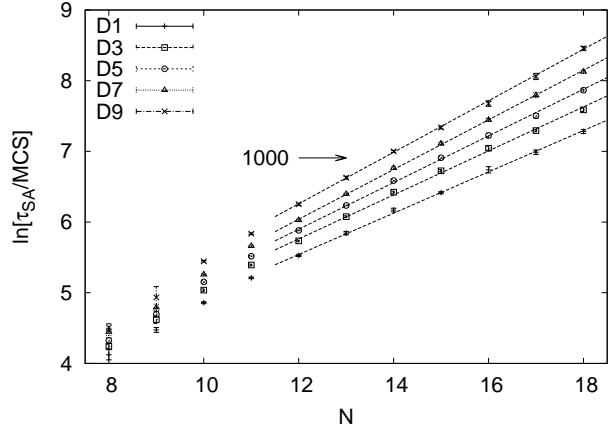}
\vspace*{0.0cm}
\caption{Run time singularity in ${\rm ln} \tau_{\rm SA}$ as given in
eq.(\ref{sa_run_time}) and for simulated annealing (SA). 
The straight lines correspond to fits 
with eq.(\ref{eq_sa_run_time_singularity}) to the various Deciles. 
Numeric values of fit parameters are found in Table 3).}
}
\label{plot_sa_run_time}
\end{figure}

~~~~~~~~~~~~~~~~~~~~~~~~~~~~~~~~~~~~~~~~~
\begin{table}[htb*]
\centering{
\begin{tabular}{|c|c|c|c|c|c|} 
\hline      
${\rm Decile}$ & $N_{\rm min}$ & $r_{\rm SA}$ & $\chi^2_{dof}$\\
\hline\hline
 1 & 12 & 0.293 ( 4 )& 0.9\\ 
 3 & 12 & 0.311 ( 6 )& 4.4\\ 
 5 & 12 & {\bf 0.330 ( 5 )}& 1.9\\ 
 7 & 12 & 0.351 ( 3 ) & 0.8\\ 
 9 & 12 & 0.364 ( 3 ) & 0.5\\ 
\hline\hline
\end{tabular}
\caption{Rate constants $r_{\rm SA}$ of the exponential run time 
singularity, see eq.(\ref{eq_sa_run_time_singularity}) in
simulated annealing (SA).
The fits have been performed with $N \ge N_{\rm min}$.
The $\chi^2_{dof}$-values of the fit are still of reasonable magnitude. 
The dependence of $r_{\rm SA}$ on the Decile is weak and D5 
denotes the median. The median rate constant is highlighted.}
}
\end{table}

There is a simple question: does quantum 
annealing with the help of quantum fluctuations 
perform differently, than classical annealing under 
the use of thermal fluctuations for a given problem set ? 
Or on the practical side: Can quantum annealing
outperform classical annealing ?
The issue here is narrowed down to the very specific  
comparison of the findings in LZ theory (QA) to the 
numerical data of simulated annealing (SA). 
The scope of the comparison is limited and can possibly
be broadened in sub-sequent work.
On the quantum side 
neither finite run-time corrections to LZ-theory 
via the excitations to 
higher energy levels, nor the corrections due 
to non-linear schedules are accounted.
On the classical side physical dynamics is approximated by 
Monte Carlo pseudo dynamics
which one may conjecture to perform faster
than genuine classical dynamics. 
We think that neither restriction is relevant here 
as efficiencies turn out to be grossly different
and likely can not be leveled by corrections.

We display in Fig. 10) our findings with respect to 
classical dynamics. The logarithmic 
search time ${\rm ln} \tau_{\rm SA}$
of eq.(\ref{sa_run_time}) in units of Monte Carlo steps
is displayed as a function of the spin number $N$ for a selected
sub-set of Deciles on the problem set. For spin numbers $N \ge 10$
up to $N=18$ we find a exponential run time singularity of the form
\be
\tau_{\rm SA}/{\rm MCS} ~=~A~e^{+r_{\rm SA} N}~~~
\label{eq_sa_run_time_singularity}
\ee
and the fitted rate constants $r_{\rm SA}$ for 
various Deciles $D1,D3,D5,D7$
and $D9$ are presented in Table 3). 
Statistical error bars
have been determined with jack-knife binning methods.
The numerical rate constant value in the median (D5) 
is $r_{\rm SA}=0.330 ( 5 )$
and turns out to be identical with
the $r_{\rm DOS}=0.329 ( 2 )$ value
of the corresponding singularity in the density of states $\Omega_1(N)$. 
There is a certain spread in the rate constant values pending on the 
Decile value, which
however is bounded and thus our problem ensemble is pure:
There is no indication for an admixture of problems, that for small Deciles
would either show polynomial, or otherwise 
different from exponential run time behavior at large Deciles. 
A similar conclusion can be obtained for the quantum case.

\begin{figure}[htb*]
\centering{
a)\includegraphics[angle=-90,width=8.5cm]{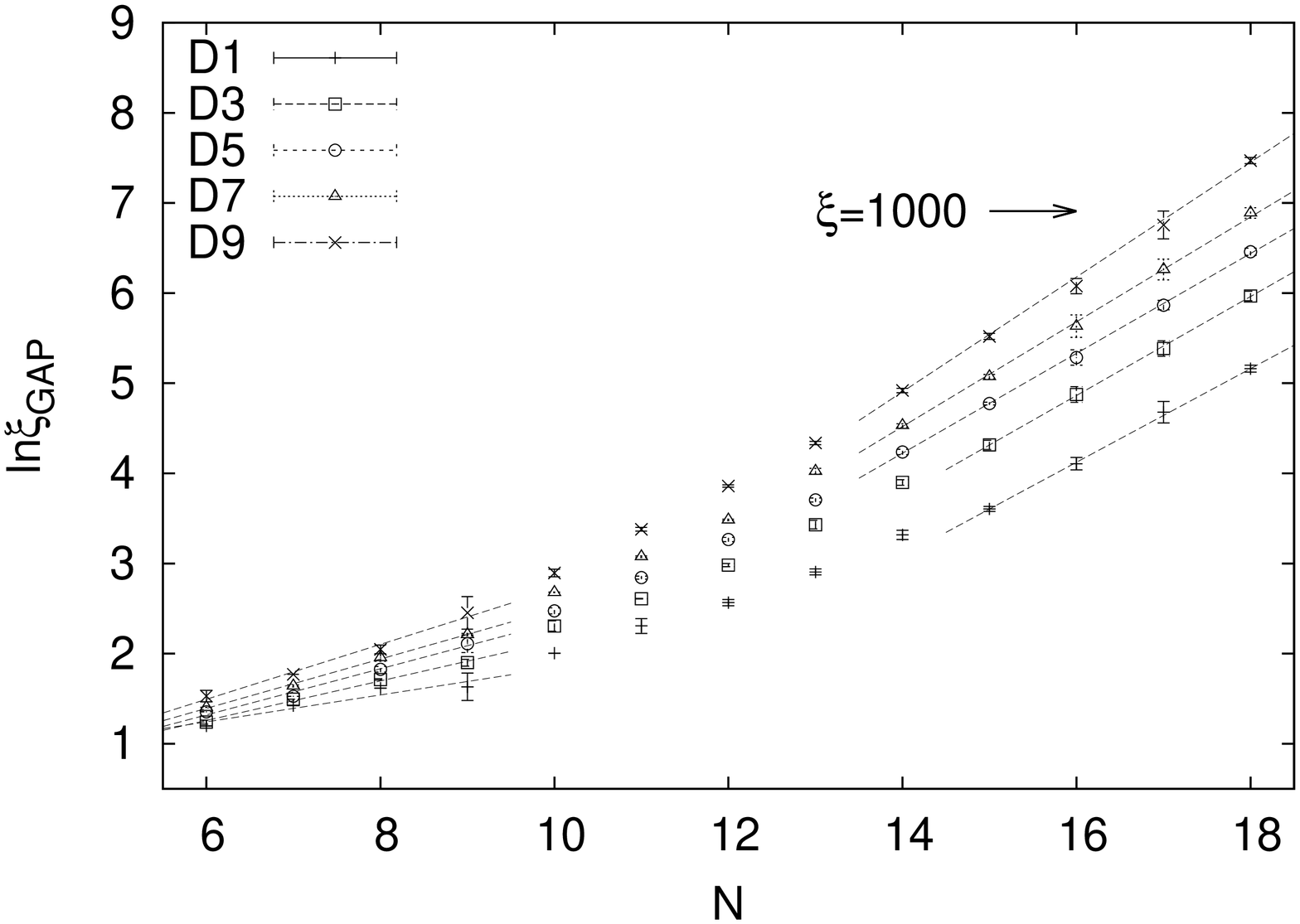} \\ b) 
\includegraphics[angle=-90,width=8.5cm]{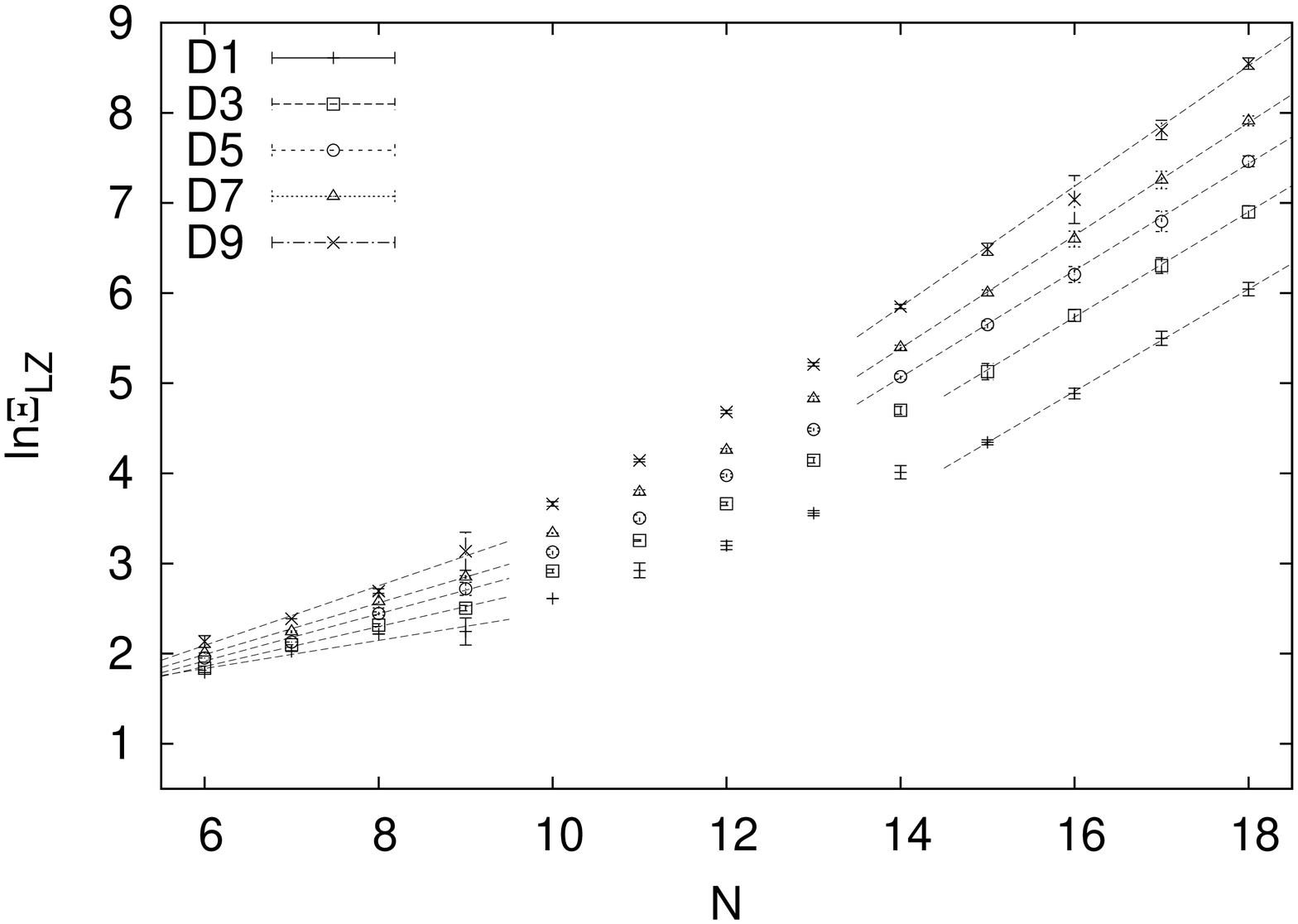}
\vspace*{0.0cm}
\caption{Logarithmic gap correlation length 
singularity ${\rm ln} \xi_{\rm GAP}$ in
in a) for a selected set of Deciles with 
D5 denoting the median as a function of $N$. 
In b) we display the root of the run 
time $\Xi_{\rm LZ}=\sqrt{\tau_{\rm QA}}$
of eq.(\ref{runtime4}) in logarithmic scale, which 
for linear schedules and 
up to regular corrections is expected to have the 
same singular behavior as ${\rm ln} \xi_{\rm GAP}.$
The straight lines in a) and b) at large $N$ correspond to fits with 
eq.(\ref{eq_gap_singularity}) and 
to eq.(\ref{eq_sqa_run_time_singularity}) to the various Deciles. 
Fitted values for the
to slopes of the lines in the figures, corresponding to rate
constants are found in Table 4). The straight lines in a) and b) at small $N$
are ment to guide the eye.}
}
\label{fig_qa_run_times}
\end{figure}
~~~~~~~~~~~~~~~~~~~~~~~~~~~~~~~~~~~~~~~~~
\begin{table}[htb*]
\centering{
\begin{tabular}{|c|c|c|c|c|c|c|c|} 
\hline      
${\rm Decile}$ & $N_{\rm min}$ & $r_{\rm GAP}$ & $\chi^2_{dof}$ & $r_{\rm LZ}$ & $\chi^2_{dof}$ \\
\hline\hline
1  & 15 & 0.519 ( 04 ) & 0.07 & 0.567 ( 8 ) & 0.12 \\ 
3  & 15 & 0.549 ( 06 ) & 0.06 & 0.584 ( 9 ) & 0.07 \\ 
5  & 14 & {\bf 0.553 ( 06 )} & 0.30 & {\bf 0.592 ( 8 )} & 0.23 \\ 
7  & 14 & 0.581 ( 12 ) & 0.80 & 0.625 ( 6 ) & 0.19 \\ 
9  & 14 & 0.636 ( 09 ) & 0.85 & 0.669 ( 8 ) & 0.28 \\ 
\hline\hline
\end{tabular}
\caption{ Fitted rate constants $r_{\rm GAP}$ (third column) 
and $r_{\rm LZ}$ (fifth column) of the singularities in 
eq.(\ref{eq_gap_singularity}) and 
to eq.(\ref{eq_sqa_run_time_singularity}) for a selected set
of deciles. The dependence on the Deciles is weak and D5 
denotes the median (displayed in bold face) in both cases.
The fits have been performed with $N \ge N_{\rm min}$ 
and yield $\chi^2_{dof}$-values
as given in columns four and six, indicating good quality fits.}
}
\end{table}

We display in Figures 11 a) and 11 b) our findings with 
respect to quantum search complexity. 
The logarithmic gap correlation length ${\rm ln}
\xi_{\rm GAP}$ is given in  Fig. 11 a) while the logarithmic 
run time root ${\rm ln} \Xi_{\rm LZ}$ 
with $\Xi_{\rm LZ}=\sqrt{\tau_{\rm QA}}$ is given in Fig. 11 b).
Again we present data for a selected set of Deciles including the median.
We observe a crossover behavior from weak singular
behavior at small $N$ turning to strong singular behavior at largest $N$.
For spin numbers $N \ge 14$
and for both observables the data 
are consistent with exponential singularities 
\be
\xi_{\rm GAP}~=~A~e^{+r_{\rm GAP} N}
\label{eq_gap_singularity}
\ee
and 
\be
\Xi_{\rm LZ}~=~A~e^{+r_{\rm LZ} N}.
\label{eq_sqa_run_time_singularity}
\ee
Exponential singular behavior in run-times
was also observed by Znidaric in \cite{Znidaric_2005} 
for 3SAT with a strength of the singularity that
at given $N$ however was weaker then the one in this paper.
The numerical rate constant values in the median 
(D5) as obtained from fits to the data
with $N \ge 14$ are $r_{\rm GAP}=0.553 ( 6 )$
and $r_{\rm LZ}=0.592 ( 8 )$ and are given in Table 4).  
The fitted values of rate constants for
other Deciles and corresponding fit quality values 
$\chi^2_{d.o.f.}$ can also be found in Table 4). 
The rate constant  $r_{\rm LZ}$ turns out to be
ten percent larger than $r_{\rm GAP}$ and all the fits
have $\chi^2_{d.o.f.}$-values of ${\cal O}(1)$ indicating good
quality data modeling.
 
Summarizing, using a conventional standard quantum 
adiabatic algorithm with a transverse field 
and a linear  time schedule we find a ground state 
quantum search complexity for the ensemble of hard 2SAT 
problems~\cite{Neuhaus_on_hard_problems} with a super hard
\begin{equation}
\tau_{\rm QA} (N)=~ {\rm const}~e^{+r_{\rm QA}N}
\end{equation}
run time singularity at $r_{\rm QA}={1.184(16)}$, that even 
exceeds the $2^N$ singularity of trivial enumerations. 
The corresponding singularity in classical, i.e. simulated 
annealing (SA) searches has 
a rate constant $r_{\rm SA}$ which only is a fraction 
$r_{\rm SA}/r_{\rm QA} \approx 0.34$.
A compute time wall of such 
magnitude for an adiabatic quantum algorithm 
designed to solve the 2SAT problem 
is theoretically interesting, but from an algorithmic point of view
can only be termed an failure. 
Finally, and recalling that our problems are 
actually in ${\cal P}$, we state the obvious 
fact that quantum annealing, as 
well as classical annealing  
are not able to
recognize a mathematical structure which otherwise and
using calculus makes the problem solvable in polynomial time.
It is unclear as to which basis for the degrees of
freedoms, or which Hamiltonian had to be chosen in 2SAT 
in order that annealing can succeed in polynomial time, or whether
that is possible at all. 



\section{Conclusion}

Within the scope of the present work we 
study the quantum search complexity within hard 2SAT
which is mapped to an ensemble of Ising models. 
We hide a unique ground state
at energy zero from an exponentially 
large number of configurations at the 
classical energy gap value ($E=1$). 
The situation corresponds to a search for an 
needle in a haystack, where the haystack 
extents over the energy one surface.
For our theory the free energy landscape 
at energy one is simple and
all configurations there are connected
by single spin flips. 
There are concurrent proposals for 
Ising models \cite{Katzgraber_02014} i.e., 
physical glasses e.g. the 3D EA glass with less trivial 
connectivity properties in vicinity of 
the ground state energy, which are conjectured to be 
candidate models for quantum speed ups over classical annealing.
However, for the specific problem set here
we find a quantum search run-time $\tau_{\rm QA}$,
that scales unfavorable to large values of $N$
with an exponential singularity.
The rate constant of the singular behavior
$r_{\rm QA}= 1.184(16)$
turns out to be larger than 
${\rm ln}2$ of trivial 
phase space enumerations !
Thus even a simple enumeration 
wins over quantum annealing for large $N$
i.e., finds ground states faster than QA. 
In addition stimulated annealing finds ground states
faster than enumeration. 
We do not expect that gradual 
changes to the algorithm: neither
alternate annealing schedules, nor 
alternate driver or problem 
Hamiltonian's can change 
this ranking as long as energies 
above and at the ground state are 
populated like a needle in a haystack.

A quantum search algorithm, that for an tractable 
theory in ${\cal  P}$ blatantly fails with compute 
times that exceed Ising enumeration provokes 
a question: Is there place for a  
loophole that could possibly restore confidence too some degree 
into the efficiency of quantum annealing ? 
For the specific case of 2SAT we note that the 
mathematical search for logical 
collisions on 2SAT's implication 
graph can actually be shaped into the form of an alternate 
quantum adiabatic algorithm. The algorithm checks for logical collisions
i.e., the connectivity of literals with anti-literals on the directed
implication graph. Whether such a quantum algorithm yields better 
efficiency is currently not known.

We have taken effort to predict the shape of success probability
distributions in quantum annealing, which turn out 
to be bimodal at mean success one half.
These probabilities are important functions as 
for annealing devices they correspond to direct 
physical observations. 
We then identify the origin of the bimodal 
property. Within LZ theory 
it is caused by a Weibull distributed 
static quantifier $\Xi_{\rm LZ}$, which is 
propagated with the help of the quadratic
Landau Zener run-time pole into a bimodal distribution. 
However, the Weibull distribution  
by itself is not predicted by theory. 
It rather characterizes the problem set. 
We also add the interesting general remark 
that quantifier distributions of the Frechet type at $k=2$
would yield constant success probability 
distribution functions, which however is not the case 
for our problem set.
We expect that a similar mechanism 
generates the bimodal successes for the EA glass
on Dwave \cite{Troyer_02013}. The different shapes
of PDF's here and there are likely caused by different 
quantifier distributions of the catastrophic type, which on Dwave
are not yet known.

We also note that our problems 
can be mapped onto the Dwave computer. 
The procedure is not elegant and will require 
problem embedding on Dwave's 
Chimera topology. 
Should the machine follow
quantum dynamics at low temperatures, then the scaling 
of the run times with $N$ ought to exhibit the failure 
of the quantum search quite clearly, as we know now that  
compute time barriers for the quantum search 
are substantially harder than for the classical search.
A failure of this kind would actually be welcome and 
it would support recent findings on the quantum nature of the 
search process \cite{Lidar_02014}.
It is however equally reasonable to expect that the
machine could solve the hard 2SAT problems with ease, in 
which case this would hint at a classical search mode on the machine, 
consistent with \cite{Smolin_02013,Smolin_02014}.
We hope to actually perform the quantum
computer experiment in the near future. 

Finally we remark that \cite{Neuhaus_on_hard_problems} contains
several KSAT theories with $K=2,...,6$ and corresponding
hard KSAT problems. For $K \ge 3$, e.g. 3SAT these can be mapped via
polynomial transformations to maximal independent 
set (MIS) \cite{VChoi_02010} with an alternate Ising
problem Hamiltonian. As the models 
at $K \ge 3$ appear even harder we expect 
even more pronounced failures of quantum 
annealing with increasing $K$ if searches are quantum.

{\bf Acknowledgment:} Calculations were performed under the 
VSR grant JJSC02 and an Institute account SLQIP00 at J\"ulich
Supercomputing Center on various computers.

\end{document}